\documentclass[aps,prd,showpacs,preprintnumbers,nofootinbib,superscriptaddress,notitlepage]{revtex4-1}

\usepackage{amsmath}
\usepackage{amssymb}
\usepackage{latexsym}
\usepackage{amsfonts}
\usepackage{epsfig}
\usepackage{psfrag}
\usepackage{graphicx}
\usepackage{wasysym}
\usepackage{ulem}
\usepackage{xcolor}
\usepackage{subfigure}

\newcommand{\Eqref}[1]{Eq.~\eqref{#1}}
\renewcommand{\vec}[1]{\mathbf{#1}}

\begin{document}

\title{Quantum reflection of photons off spatio-temporal electromagnetic field inhomogeneities}

\author{Holger Gies}
\author{Felix Karbstein}
\author{Nico Seegert}
\affiliation{Theoretisch-Physikalisches Institut, Abbe Center of Photonics,
Friedrich-Schiller-Universit\"at Jena, Max-Wien-Platz 1, D-07743 Jena, Germany}
\affiliation{Helmholtz-Institut Jena, Fr\"obelstieg 3, D-07743 Jena, Germany}

\begin{abstract}

  We reconsider the recently proposed nonlinear QED effect of
  quantum reflection of photons off an inhomogeneous strong-field
  region. We present new results for strong fields varying both in
  space and time. While such
  configurations can give rise to new effects such as frequency
  mixing, estimated reflection rates based on previous one-dimensional
  studies are corroborated. On a conceptual level, we critically
  re-examine the validity regime of the conventional
  locally-constant-field approximation and identify kinematic
  configurations which can be treated reliably. Our results further
  underline the discovery potential of quantum reflection as a new
  signature of the nonlinearity of the quantum vacuum.

\end{abstract}

\date{\today}


\maketitle

\section{Introduction}

The vacuum of quantum electrodynamics (QED) can be considered as
permeated by virtual electron-positron fluctuations which probe
typical distances of the order of the electron Compton wavelength
$\lambda_c=1/m$ and exist for typical time scales of the order of the
Compton time $\tau_c=1/m$.  Here $m$ is the electron mass, and we use
units where $\hbar=c=1$.
As electromagnetic fields couple to charges, these virtual charged
particle fluctuations can induce effective nonlinear couplings between
electromagnetic fields
\cite{Heisenberg:1935qt,Weisskopf,Schwinger:1951nm}, and affect the
propagation of photons.  So far, the pure electromagnetic nonlinearity
of the quantum vacuum though subject to high-energy experiments
\cite{Akhmadaliev:1998zz,Akhmadaliev:2001ik} has not been directly
verified on macroscopic scales.

In a recent paper \cite{Gies:2013yxa}, we proposed quantum reflection
as a new signature of quantum vacuum nonlinearity in the presence of
strong electromagnetic fields. This phenomenon complements commonly
studied nonlinear vacuum effects such as direct light-by-light
scattering \cite{Euler:1935zz,Karplus:1950zz}, vacuum magnetic
birefringence \cite{Toll:1952,Baier,BialynickaBirula:1970vy}, photon
splitting \cite{Adler:1971wn}, and spontaneous vacuum decay in terms
of Schwinger pair-production in electric fields
\cite{Sauter:1931zz,Heisenberg:1935qt,Schwinger:1951nm}; for recent
reviews, see
\cite{Dittrich:2000zu,Marklund:2008gj,Dunne:2008kc,DiPiazza:2011tq,Battesti:2012hf}.

Many proposals to verify quantum vacuum nonlinearities under
controlled laboratory conditions rely on a pump-probe scheme, where a
well-controlled probe photon beam traverses a region of space subject
to a strong electromagnetic field (``pump'').  Typical examples for an
experimental realization of such a scheme are experiments intended to
verify vacuum birefringence in macroscopic magnetic fields
\cite{Cantatore:2008zz,Berceau:2011zz}.  An analogous scheme to be
realized with the aid of high-intensity lasers has been proposed in
\cite{Heinzl:2006xc}, see also \cite{DiPiazza:2006pr,Heinzl:2008an,Dinu:2013gaa}.
Alternative concepts suggest the use of time-varying fields and high-precision interferometry \cite{Zavattini:2008cr,Dobrich:2009kd,Grote:2014hja}.

With regard to the phenomenon of quantum reflection, we emphasize the viewpoint that the quantum vacuum subject to strong electromagnetic fields can be interpreted as constituting an effective attractive potential for the traversing probe photons (cf. \cite{Gies:2013yxa}).
Even though the effective potential is attractive, the traversing photons can experience reflection due to the phenomenon of above-barrier scattering \cite{QR1}, which -- as it has no analogue in classical physics -- is also called quantum reflection \cite{QR2}.
In comparison to standard birefringence set-ups, where the induced quantum-vacuum signature has to be isolated from a large background, our proposal of quantum reflection inherently allows for a clear separation between signal and background,
as the signature of quantum reflection amounts to reflected photons in the field free region, facilitating the use of single-photon detection techniques. Our ultimate goal is a realistic all-optical pump probe set-up to verify quantum vacuum nonlinearity. 

Whereas many vacuum phenomena exist in both homogeneous and
inhomogeneous pump field configurations, let us emphasize that quantum
reflection manifestly requires spatially inhomogeneous pump field
profiles.  Other optical signatures of quantum vacuum nonlinearities that
require inhomogeneous pump profiles are those based on interference
effects \cite{King:2013am,Tommasini:2010fb,Hatsagortsyan:2011},
photon-photon scattering in the form of laser-pulse collisions
\cite{King:2012aw} and the specific laser photon merging scenario
mediated by an effective four-photon, i.e., below hexagon order,
interaction considered in \cite{Gies:2014jia}.

In \cite{Gies:2013yxa} we have introduced and exemplified quantum
reflection in the case of a static one-dimensional and purely magnetic
field inhomogeneity as a prospective and promising candidate for an
all-optical probe of quantum vacuum nonlinearity.  The present
follow-up study is devoted to the discussion of quantum reflection in
more general spatio-temporal electromagnetic field
inhomogeneities. Again we resort to the locally constant field
approximation.  As will be detailed in the main part of this paper,
all basic effects encountered and discussed in \cite{Gies:2013yxa}
will persist.  Moreover, additional effects occur upon the inclusion
of temporally varying field inhomogeneities as well as more general
electromagnetic field alignments, involving, e.g., energy exchange
processes between probe photons and the field
inhomogeneity. Such frequency mixing processes have already been
studied in various contexts and suggested as a signature of
nonlinearities of QED \cite{Ding:1989zz,Rozanov:1993aa,Moulin:2002ya,Lundstrom:2005za} and
beyond \cite{Dobrich:2010hi}.

Our paper is organized as follows: Section~\ref{seq:Gen} reviews our
approach to tackle the scenario of probe photons experiencing quantum
reflection in electromagnetic fields introduced in
\cite{Gies:2013yxa}, putting special emphasis on the consideration of
higher-dimensional field inhomogeneities.  In Secs.~\ref{seq:Mag} and
\ref{seq:Cross} we discuss explicitly various types of inhomogeneities
and field configurations accessible and treatable within the framework
of the locally constant field approximation.  These examples encompass
both purely magnetic field inhomogeneities as well as configurations
involving orthogonal electric and magnetic fields (crossed fields).
Section~\ref{seq:Exp} gives estimates of the prospective number of
quantum reflected photons attainable with present and near future
laser facilities.  We end with conclusions in Sec.~\ref{seq:Con}.

\section{Quantum Reflection} \label{seq:QR}

\subsection{Towards quantum reflection in spatio-temporal electromagnetic field inhomogeneities} \label{seq:Gen}

The starting point of our analysis is the effective theory of photon propagation in an external electromagnetic field (cf., e.g. \cite{Dittrich:2000zu}). The Lagrangian is given by
\begin{equation}
\mathcal{L}[A]= -\frac{1}{4} F_{\mu\nu} F^{\mu\nu}- \frac{1}{2}\int_{x'} A_\mu(x) \Pi^{\mu\nu}(x,x'|{\cal A}) A_\nu(x')\,, \label{eq:calL}
\end{equation}
where $\Pi^{\mu\nu}(x,x'|{\cal A})$ denotes the photon polarization tensor in the presence of the external field with four-potential ${\cal A}_\mu(x)$,
accounting for the vacuum fluctuations of the underlying theory.
The field-strength tensor of the propagating photon $A_\mu$ is denoted by $F_{\mu \nu}$, and $x$ is a spatio-temporal four-vector.
Our metric convention is $g_{\mu\nu}={\rm diag}(-,+,+,+)$, such that the momentum four-vector squared reads $k^2=\vec{k}^2-\omega^2$.
Moreover, our conventions for the Fourier transform to momentum space are
$\Pi^{\mu\nu}(x,x')=\int_k\int_{k'}{\rm e}^{-ikx}\,\Pi^{\mu\nu}(k,k')\,{\rm e}^{-ik'x'}$ and $A_\mu(x)=\int_{k}\,{\rm e}^{ixk}A_\mu(k)$.

The equation of motion for probe photons is inferred straightforwardly from \Eqref{eq:calL}. In momentum space, it reads
\begin{equation}
 \left(k^2 g^{\mu \nu} - k^\mu k^\nu \right) A_\nu (k) = -\int_{k'}\!\tilde\Pi^{\mu\nu}(-k,k'|{\cal A})A_\nu(k')\,, \label{eq:EOM}
\end{equation}
where $\tilde\Pi^{\mu\nu}(k,k'|{\cal A})\equiv[\Pi^{\mu\nu}(k,k'|{\cal A})+\Pi^{\nu\mu}(k',k|{\cal A})]/2$ is the symmetrized polarization tensor.

As detailed in \cite{Gies:2013yxa}, \Eqref{eq:EOM} can be directly interpreted in terms of the scenario of quantum reflection, involving incident probe photons $A_\nu^{\rm in}$,
an inhomogeneous pump field profile ${\cal A}_\nu$ and induced photons $A_\nu^{\rm out}$:
Neglecting the backreaction of the induced photons on the incident probe photons as well as on the pump field, we can identify the photon field on the right-hand side of \Eqref{eq:EOM} with $A_\nu^{\rm in}$ and that on the left-hand side with $A_\nu^{\rm out}$.
The interaction with ${\cal A}_\nu$ is encoded in the photon polarization tensor.
Correspondingly, the right-hand side of \Eqref{eq:EOM} can be considered as a source term for outgoing reflected photons resulting from an interaction of $A_\nu^{\rm in}$ with ${\cal A}_\nu$ mediated by virtual electron-positron fluctuations.

Here, we encounter two problems:
first, \Eqref{eq:EOM} is a tensor equation of rather involved structure, which is difficult to solve in closed form for general kinematic situations and electromagnetic field configurations.
Second and conceptually even worse, the polarization tensor is not known in an explicit form for arbitrary field configurations $\mathcal{A}$.
Even though estimates of the polarization tensor for all-optical setups can be deduced from constant-field versions of the polarization tensor, it may not be straightforward to
construct approximations of $\Pi^{\mu\nu}$ that satisfy the Ward identity imposed by gauge invariance.
In the following, we show that both problems can be solved simultaneously, if two assumptions hold:
$(a)$ First we assume that the tensorial quantities in \Eqref{eq:EOM} can be represented as follows,
\begin{equation}
 \left(k^2 g^{\mu \nu} - k^\mu k^\nu \right)=k^2\sum_p (P'_p){}^{\mu\nu} \quad\quad \text{and} \quad\quad
 \tilde\Pi^{\mu\nu}(-k,k'|{\cal A})=\sum_{p} \tilde\Pi_p(-k,k'|{\cal A})\,(P_p)^{\mu\nu} + Q^{\mu\nu}, \label{eq:decompinProjs}
\end{equation}
with scalar coefficients $\tilde \Pi_p$.
Here $(P'_p)^{\mu\nu}$ and $(P_p)^{\mu\nu}$ correspond to two -- {\it a priori} unrelated -- sets of projectors fulfilling $(P_pP_q)^{\mu\nu}=\delta_{p,q}(P_p)^{\mu\nu}$, $(P_p)^\mu_{\ \mu}=1$ and similarly $(P'_pP'_q)^{\mu\nu}=\delta_{p,q}(P'_p)^{\mu\nu}$, $(P'_p)^\mu_{\ \mu}=1$.
In the first expression in \Eqref{eq:decompinProjs} an overall factor of $k^2$ can be factored out, and the projectors ${P'}_p^{\mu\nu}$ correspond to projectors onto photon polarization modes $p$ to be specified in Secs.~\ref{seq:Mag} and \ref{seq:Cross} below.
For the polarization tensor we explicitly accounted for a residual tensorial contribution $Q^{\mu\nu}$ which cannot be expressed in terms of the projectors $(P_p)^{\mu\nu}$, and does not necessarily fulfill $(P_pQ)^{\mu\nu}=0$ for all $p$.
Note that such contribution is, e.g., present in the polarization tensor for generic constant electromagnetic fields; cf. \cite{Dittrich:2000zu}.
To keep notations simple, we have omitted any explicit reference to the momentum as well as ${\cal A}_\mu$ dependencies for both sets of projectors and for $Q^{\mu\nu}$.

$(b)$ Second we assume that there is at least a single {\it global projector} $(\tilde P_p)^{\mu\nu}$ which fulfills $(\tilde P_p)^{\mu\nu}\equiv(P_p)^{\mu\nu}\equiv(P'_p)^{\mu\nu}$ and $(\tilde P_pQ)^{\mu\nu}\equiv0$.
Taking into account the above assumptions, a contraction of \Eqref{eq:EOM} with this particular projector results in
\begin{equation}
 k^2\,A^{\rm out}_p(k) = -\int_{k'} \tilde\Pi_p(-k,k'|{\cal A})\, A^{\rm in}_p(k')\,, \label{eq:EOMs}
\end{equation}
where have we introduced photons $A_p^\mu=\tilde P_p^{\mu\nu}A_\nu$  polarized in mode $p$. As the equation of motion loses any nontrivial Lorentz index structure, we have dropped the trivial Lorentz indices of the photon fields, $A_{p,\mu}(k)\to A_{p}(k)$.
The resulting scalar equation~\eqref{eq:EOMs} can be solved straightforwardly with Green's function methods.
Below, we will show that assumptions $(a)$ and $(b)$ invoked above are
in fact satisfied in several cases of physical interest. Actually, all
our subsequent considerations and insights will be based thereon.

Let us emphasize that no analytical results are available for the photon polarization tensor in generic inhomogeneous fields.
However, the polarization tensor is known explicitly at one-loop accuracy for several special background fields that can be divided into two classes:
homogeneous electric and/or magnetic fields
\cite{BatShab,narozhnyi:1968,ritus:1972,Tsai:1974fa,Tsai:1974ap,Urrutia:1977xb,Schubert:2000yt}
(cf. also \cite{Dittrich:2000zu,Karbstein:2013ufa} and references
therein), and generic plane wave backgrounds
\cite{Baier:1975ff,Becker:1974en}; see \cite{Meuren:2013oya} for a
more recent derivation and an alternative representation, and
\cite{Gies:2014jia} for a novel systematic expansion especially suited
for all-optical experimental scenarios. Numerical results for
inhomogeneous magnetic backgrounds are available from worldline Monte
Carlo simulations \cite{Gies:2011he}.

In \cite{Gies:2013yxa} we have devised a strategy to nevertheless gain analytical insights into the photon polarization tensor in the presence of inhomogeneous backgrounds,
more specifically for inhomogeneous field configurations that may locally be approximated by a constant:
As already mentioned in the introduction, virtual electron-positron fluctuations probe typical distances of ${\cal O}(\lambda_c)$ and exist for typical time scales of ${\cal O}(\tau_c)$. Hence, for inhomogeneities with typical scales of spatial and temporal variations $w$ and $\tau$ much larger than the Compton wavelength and the Compton time of the virtual charged particles, i.e., $w\gg\lambda_c$ and $\tau\gg\tau_c$,
using the constant field expressions locally is well justified.
In \cite{Gies:2013yxa} we exemplified our procedure to built in the desired inhomogeneous field profile {\it a posteriori} for the special case of a static one-dimensional
purely magnetic field inhomogeneity starting from the photon polarization tensor in the corresponding homogeneous background.
This strategy can be straightforwardly adapted to higher-dimensional field inhomogeneities.
Here we exclusively focus on inhomogeneities to be characterized by a single field amplitude profile ${\cal E}(x)$.
Aiming at the polarization tensor in the presence of a particularly oriented and shaped electromagnetic field inhomogeneity, this suggests the following two-step procedure:
First, we identify an explicitly known photon polarization tensor with the desired field alignments.
Second, we incorporate the amplitude profile of the inhomogeneity by the strategy devised in \cite{Gies:2013yxa}.
Here we limit ourselves to homogeneous background fields as starting point; starting from plane wave backgrounds would require minimal modifications (cf. also the last paragraph of Sec.~\ref{seq:c2+1} below).

As a consequence of Furry's theorem (charge conjugation symmetry of
QED), the resulting photon polarization tensor in position space is
even in the combined parameter $e{\cal E}(x)$.  A corresponding
perturbative expansion can be compactly represented in momentum space
as (cf. Sec.~III. of \cite{Gies:2013yxa})
\begin{equation} 
 \tilde\Pi^{\mu\nu}(k,k'|{\cal E})=\sum_{\ell=0}^\infty \pi^{\mu\nu}_{(2\ell)}(k,k')\int_{x'}{\rm e}^{i(k+k')x'}\,[e{\cal E}(x')]^{2\ell}\,, \label{eq:SumPiinh}
\end{equation}
where we introduced the tensorial quantity
\begin{equation}
 \pi^{\mu\nu}_{(2\ell)}(k,k')=\frac{1}{2}\Bigl[\Pi^{\mu\nu}_{(2\ell)}(k) + \Pi^{\mu\nu}_{(2\ell)}(k')\Bigr], \label{eq:pi}
\end{equation}
which is completely independent of the background field amplitude.
The expansion coefficients $\Pi^{\mu\nu}_{(2\ell)}$ can be read off from a Taylor expansion of the respective photon polarization tensor for ${\cal E}=const.$ about $e{\cal E}=0$. 

We emphasize that the polarization tensor~\eqref{eq:SumPiinh} constructed along these lines may violate the Ward identity, $k_\mu\Pi^{\mu\nu}(k,k')=0=\Pi^{\mu\nu}(k,k')k'_\nu$.
This is more obvious in position space, where the Ward identity reads $\partial_\mu \Pi^{\mu\nu}(x,x'|\mathcal{E})=0=\partial_\nu'\Pi^{\mu\nu}(x,x'|\mathcal{E})$  holding for any background-field $\mathcal{E}$;
if now $\Pi^{\mu\nu}(x,x'|\mathcal{E})$ has been computed for constant $\mathcal{E}$, the naive insertion of an inhomogeneous field $\mathcal{E}(x)$ into $\Pi^{\mu\nu}(x,x'|\mathcal{E})$ can generically be expected to be in conflict with the Ward identity.
This is, because the position space derivatives will also hit the position space dependence of the background field $\mathcal{E}(x)$.

The important point now is that the assumptions $(a)$ and $(b)$ above in fact amount to a restriction on tensorial components of \Eqref{eq:SumPiinh} for which the Ward identify is fully satisfied also for an inhomogeneous $\mathcal{E}$:
For the explicitly known expressions of the polarization tensor in homogeneous and plane wave backgrounds serving as starting point in our locally-constant-field approximation, the tensor structure can always -- at least partially -- be written in terms of projectors onto photon polarization modes.
Of course the outgoing photon projectors $(P'_p)^{\mu\nu}$ in \Eqref{eq:decompinProjs} can be chosen accordingly, and the transversal structure $(k^2 g^{\mu \nu} - k^\mu k^\nu)$ be decomposed in terms of these projectors.
For \Eqref{eq:SumPiinh} to be compatible with the assumptions $(a)$ and $(b)$ above, we have to single out a projector which constitutes a global projector of the polarization tensor~\eqref{eq:SumPiinh},
i.e., fulfills $(\tilde P_p)^{\mu\nu}\equiv(P_p)^{\mu\nu}(k)=(P_p)^{\mu\nu}(k')$.
This in turn implies that the assumptions $(a)$ and $(b)$ together with \Eqref{eq:SumPiinh} ensure the Ward identity to remain intact for the tensorial structure of the polarization tensor~\eqref{eq:SumPiinh} projected out by the global projector $(\tilde P_p)^{\mu\nu}$,
as then of course $k_\mu(\tilde P_p)^{\mu\nu}=(\tilde P_p)^{\mu\nu}k'_\nu=0$ holds exactly.

As in \cite{Gies:2013yxa}, we limit ourselves to incident probe photons with wave vector in the ${\rm x}$-${\rm y}$ plane  in the present paper.
Moreover, the incident probe photons will always be assumed to propagate on the light cone and to be described by a positive-energy plane wave,
$A^{\rm in}_p(x)=a(\omega_{\rm in})\,{\rm e}^{i\omega_{\rm in}({\rm x}\cos\beta+{\rm y}\sin\beta-t)}$ of energy $\omega_{\rm in}>0$ and amplitude $a(\omega_{\rm in})$, whose propagation direction is controlled by the choice of the angle $\beta\in(-\pi\ldots\pi]$.

Aiming at the detection of induced outgoing photons in the field free
region at distances far away from the inhomogeneity, we transform
\Eqref{eq:EOMs} with the appropriate boundary conditions to position
space,
\begin{equation}
 \square A^{\rm out}_p(x) = \int_{x'} \tilde\Pi_p(x,x'|{\cal A})\, A^{\rm in}_p(x')\equiv j_p(x)\,. \label{eq:EOMs_poss}
\end{equation}
The induced current $j_p(x)$ in \Eqref{eq:EOMs_poss} is nonzero only where the background field deviates from zero, i.e., within the spacetime region where the electromagnetic field inhomogeneity is localized, and falls off rapidly to zero outside.
This implies that the incident probe photons can only give rise to induced outgoing photons in the region where they experience a nonvanishing background electromagnetic field.
It is then convenient to make use of a mixed position-momentum space representation of the photon polarization tensor,
i.e., perform a partial Fourier transform with respect to its second argument, and write
\begin{equation}
 j_p(x)  = \tilde\Pi_p(x,k_{\rm in}|{\cal A})\,a(\omega_{\rm in})\,, \label{eq:EOMs_poss1}
\end{equation}
with $k_{\rm in}^\mu=\omega_{\rm in}(1,\cos\beta,\sin\beta,0)$, fulfilling $k_{\rm in}^2=0$.
As $j_p(x)$ is nonzero only in the spacetime region where the electromagnetic field inhomogeneity is localized,
we can extend the integration ranges to $\pm\infty$ when determining $ A^{\rm out}_p(x)$ in terms of a Green's function integral,
\begin{equation}
 A^{\rm out}_p(x)=\int_{x'} G^{\rm R}_+(x,x')\,j_p(x')\,. \label{eq:Asim}
\end{equation}
Here $G^R_+$ denotes the positive energy branch\footnote{With regard to the Fock space representation of the photon field in position space $|A_p(x)\rangle=\int \frac{{\rm d}^3k}{(2\pi)^3}\,\frac{1}{\sqrt{2k^0}}\,{\rm e}^{ikx} a^\dag_{\vec{k},p}|0\rangle$, where $k^0=\sqrt{\vec{k}^2}$, only positive energy waves amount to propagating real photons.}
of the retarded Green's function $G^R=G^R_++G^R_-$ for the D'Alembert operator $\square=\partial^\mu\partial_\mu$,
which is determined straightforwardly in momentum space, where its defining equation reads
\begin{equation}
 -\bigl[\vec{k}^2-(\omega+i\epsilon)^2\bigr]G^{\rm R}(k,k')=(2\pi)^4\,\delta(k+k')\,.
\end{equation}
A useful integral representation of this Green's function is
\begin{equation}
 G^{\rm R}(x,x')=\Theta(t-t')\int\frac{{\rm d}^3k}{(2\pi)^3}\,\frac{1}{2i\sqrt{\vec{k}^2}}\,\Bigl({\rm e}^{i k_+(x-x')}-{\rm e}^{i k_-(x-x')}\Bigr), \label{eq:GR}
\end{equation}
with $k^\mu_\pm\equiv(\pm\sqrt{\vec{k}^2},\vec{k})$ and $k_\pm^2=0$, encompassing both positive and negative energy waves propagating on the light cone.
Causality is ensured by the overall Heaviside function, nullifying $G^{\rm R}(x,x')$ for $t-t'<0$.
In a sense this Heaviside function also provides the time arrow, and thus a means to distinguish between positive and negative energies.
Correspondingly, we have
\begin{equation}
 G^{\rm R}_\pm(x,x')=\pm\Theta(t-t')\int\frac{{\rm d}^3k}{(2\pi)^3}\,\frac{1}{2i\sqrt{\vec{k}^2}}\,{\rm e}^{i k_\pm(x-x')}. \label{eq:GRpm}
\end{equation}

In our calculation we will always assume $t-t'>0$ to be fulfilled, such that the detection takes place at times $t$ after the interaction, but nevertheless extend the $t'$ integration up to $+\infty$.
For inhomogeneities localized in time this is clearly justified for the reasons given above.\footnote{
Let us briefly comment on a subtlety associated with field inhomogeneities infinitely extended in time.
Namely, we argue that also in this case we can formally extend the $t'$ integration up to $+\infty$: 
Our calculation, assuming a continuous inflow of plane-wave probe photons actually is for a steady state.
Hence, we do not expect the induced photon fields to exhibit any explicit time dependence apart from a trivial time dependence in the exponential.
Other explicit references to time scales in intermediate steps of the calculation should drop out.
The corresponding scattering rates mediating incident probe photons into induced photons involve a modulus squared and are then rendered time independent, as expected for a steady state (cf. below).
Moreover, for a given incident photon, the interaction with the inhomogeneity is always limited to that section of its trajectory traversing the inhomogeneity (cf. above). 
Hence, also in case of a field inhomogeneity infinitely extended in time the detection takes place well after the actual interaction, and can formally be assumed to happen at asymptotic times.}

Plugging the integral representation~\eqref{eq:GRpm} of $G_+^R$ into \Eqref{eq:Asim} and making use of $\int_{x'} {\rm e}^{-i k_+ x'}\,\tilde\Pi_p(x',k_{\rm in}|{\cal A}) = \tilde\Pi_p(-k_+,k_{\rm in}|{\cal A})$, we finally obtain
\begin{equation}
 A^{\rm out}_p(x)= a(\omega_{\rm in})\int\frac{{\rm d}^3k}{(2\pi)^3}\,{\rm e}^{i k_+ x}\,\frac{\tilde\Pi_p(-k_+,k_{\rm in}|{\cal A})}{2i\sqrt{\vec{k}^2}}\,, \label{eq:Aindout}
\end{equation}
i.e., the induced photon field can be written in terms of a Fourier integral of the photon polarization tensor in momentum space $\tilde\Pi_p(k,k'|{\cal A})$ accounting for a given field inhomogeneity ${\cal A}(x)$.

As both in- and outgoing four-momenta fulfill $k_{\rm in}^2=k_+^2=0$, the photon polarization tensor only has to be evaluated on the light cone.
In this limit it is of a particularly simple form.
We emphasize that \Eqref{eq:Aindout} accounts for all induced positive energy photon modes. An induced contribution is to be considered as physical at a given far-field position if it amounts to an outgoing wave originating from the inhomogeneity.

When adapted to the particular kinematics in \Eqref{eq:Aindout} and contracted with the respective global projector, \Eqref{eq:SumPiinh} reads
\begin{equation}
 \tilde\Pi_p(-k_+,k_{\rm in}|{\cal A})=\sum_{\ell=1}^\infty\pi_{p,(2\ell)}(-k_+,k_{\rm in})\int_{x'}{\rm e}^{i(k_{\rm in}-k_+)x'}\,[e{\cal E}(x')]^{2\ell}\,, \label{eq:Pip_speckin}
\end{equation}
where we omitted the zero field ($n=0$) contribution as it vanishes on the light cone.
If an electromagnetic field inhomogeneity characterized by ${\cal E}(x)$ does not depend on a given coordinate, \Eqref{eq:Pip_speckin} gives rise to a $\delta$ function ensuring the conservation of the corresponding momentum component.
This in turn can be employed to reduce the dimension of the integral to be performed in \Eqref{eq:Aindout}.
If, e.g., the inhomogeneity depends on the ${\rm x}$ coordinate only, i.e., ${\cal E}(x)={\cal E}({\rm x})$, we show that the result for a static one-dimensional inhomogeneity derived in \cite{Gies:2013yxa} is reproduced.
In this limit \Eqref{eq:Pip_speckin} becomes 
\begin{equation}
 \tilde\Pi_p(-k_+,k_{\rm in}|{\cal A})=(2\pi)^3\,\delta(k_{\rm y}-\omega_{\rm in}\sin\beta)\,\delta(k_{\rm z})\,\delta(\sqrt{\vec{k}^2}-\omega_{\rm in})\,
\tilde\Pi_p^{1{\rm dim}}(-k_{\rm x},\omega_{\rm in}\cos\beta|{\cal A}) \,,
\end{equation}
with
\begin{equation}
 \tilde\Pi_p^{1{\rm dim}}(-k_{\rm x},\omega_{\rm in}\cos\beta|{\cal A})=\sum_{\ell=1}^\infty\pi_{p,(2\ell)}(-k_+,k_{\rm in})\big|_{k_{\rm y}=\omega_{\rm in}\sin\beta,k_{\rm z}=0}\int{\rm dx'}\,{\rm e}^{i(\omega_{\rm in}\cos\beta-k_{\rm x}){\rm x}'}\,[e{\cal E}({\rm x}')]^{2\ell}\,, \label{eq:Pi1dim}
\end{equation}
where we only account for the non-conserved $\rm x$ momentum components in the argument of $\tilde\Pi_p^{1{\rm dim}}$. Taking into account the $\delta$ functions for the $y$ and $z$ momentum components, we can rewrite the $\delta$ function implementing energy conservation as 
$\delta(\sqrt{\vec{k}^2}-\omega_{\rm in}) \to \frac{1}{|\cos\beta|}\sum_{l=\pm} \delta(k_{\rm x}-l\omega_{\rm in}\cos\beta)$.
Correspondingly, upon performing the integration over three-momenta, \Eqref{eq:Aindout} decomposes into two contributions,
\begin{equation}
 A^{\rm out}_p(x)=A^{\rm in}_p(x)
 \biggl\{ \frac{\tilde\Pi_p^{1{\rm dim}}(-\omega_{\rm in}\cos\beta,\omega_{\rm in}\cos\beta|{\cal A})}{2i\omega_{\rm in}|\cos\beta|}
 +{\rm e}^{-i2\omega_{\rm in}{\rm x}\cos\beta}\,\frac{\tilde\Pi_p^{1{\rm dim}}(\omega_{\rm in}\cos\beta,\omega_{\rm in}\cos\beta|{\cal A})}{2i\omega_{\rm in}|\cos\beta|}\biggr\} \,, \label{eq:Aout_1d_Pi}
\end{equation}
with $A^{\rm in}_p(x)=a(\omega_{\rm in})\,{\rm e}^{i\omega_{\rm in}({\rm x}\cos\beta+{\rm y}\sin\beta-t)}$.
The first expression in the curly braces describes induced photons
transmitted in forward direction with unmodified momentum components.
The second expression amounts to the reflected part of the induced
photon field, with all momentum components apart from the ${\rm x}$
momentum component conserved.  In the latter case the ${\rm x}$
momentum component is reversed, i.e., a momentum of magnitude
$2\omega_{\rm in}|\cos\beta|$ has been absorbed from the
electromagnetic field inhomogeneity.
The associated reflection coefficient is defined as the modulus squared of the reflected wave normalized by the incident wave amplitude and reads
\begin{equation}
 R_p=\biggl|\frac{\tilde\Pi_p^{1{\rm dim}}(\omega_{\rm in}\cos\beta,\omega_{\rm in}\cos\beta|{\cal A})}{2\omega_{\rm in}\cos\beta}\biggr|^2\,, \label{eq:R1d}
\end{equation}
which agrees with the expression given in Eq.~(17) of \cite{Gies:2013yxa}.

In the following, we go beyond static one-dimensional field inhomogeneities and
discuss two generic classes of inhomogeneities, characterized by a single field amplitude profile ${\cal E}(x)$:
First, purely magnetic field inhomogeneities for which $\vec{B}(x)={\cal E}(x)\hat{\vec{B}}$, with the magnetic field pointing into a fixed direction $\hat{\vec B}$.
Second, electromagnetic field inhomogeneities featuring both electric $\vec{E}(x)={\cal E}(x)\hat{\vec{E}}$ and magnetic $\vec{B}(x)={\cal E}(x)\hat{\vec{B}}$ field components constrained by $\hat{\vec E}\perp\hat{\vec B}$,
with fixed unit vectors $\hat{\vec E}$ and $\hat{\vec B}$.

\subsection{Purely magnetic background fields} \label{seq:Mag}

In the purely magnetic field case $\hat{\vec{B}}$ provides a global spatial reference direction, with respect to which vectors can be decomposed into parallel and perpendicular components. In particular,
\begin{equation}
 k^\mu=k^\mu_\parallel + k^\mu_\perp\,,\quad k_\parallel^\mu=(\omega,\vec{k}_\parallel)\,,\quad k_\perp^\mu=(0,\vec{k}_\perp)\,,
\end{equation}
with $\vec{k}_\parallel=(\vec{k}\cdot\hat{\vec{B}})\hat{\vec{B}}$ and $\vec{k}_\perp=\vec{k}-\vec{k}_\parallel$.
In the same way tensors can be decomposed, e.g., $g^{\mu\nu}=g_\parallel^{\mu\nu}+g_\perp^{\mu\nu}$.
As detailed in the previous section, to determine the induced photon field~\eqref{eq:Aindout} we only need the photon polarization tensor on the light cone.
In this particular limit and for $\vec{k} \nparallel \vec{B}$, which is implicitly assumed here (for subtleties associated with the special alignment $\vec{k} \parallel \vec{B}$ we refer to \cite{Karbstein:2013ufa}), the tensor structure of the respective photon polarization tensor is spanned by just two projectors,

\begin{equation}
 P^{\mu\nu}_\parallel(k)=g^{\mu\nu}_\parallel-\frac{k_\parallel^\mu k_\parallel^\nu}{k_\parallel^2}\,,\quad P^{\mu\nu}_\perp(k)=g^{\mu\nu}_\perp-\frac{k_\perp^\mu k_\perp^\nu}{k_\perp^2}\, . \label{eq:Proj}
\end{equation}
They project onto photon modes polarized parallel and perpendicular to the plane spanned by $\vec{B}$ and $\vec{k}$, respectively.

The main effort in adopting the strategy outlined above is to identify inhomogeneous field configurations compatible with the existence of a global projector $(\tilde P_p)^{\mu\nu}$.
We emphasize that that even though the polarization tensor is finally only evaluated on the light cone [cf. \Eqref{eq:Aindout}],
the requirement for a global projector is that it remains invariant both on and off the light cone.
Taking into account \Eqref{eq:SumPiinh} the induced photon field is then inferred straightforwardly from \Eqref{eq:Aindout}.

Each of the projectors in \Eqref{eq:Proj} depends on two independent four-momentum components.
This implies that in order to allow for the definition of an invariant global projector as required by our approach, at least two four-momentum components have to remain unaffected by the inhomogeneity.
The momentum components to be affected can be inferred from the coordinates the amplitude profile ${\cal E}$ depends on,
e.g., an inhomogeneity described by the amplitude profile ${\cal E}({\rm x},t)$ can impact photon energies and momenta along ${\vec e}_{\rm x}$.
Correspondingly, in the magnetic field case we can study inhomogeneities varying at most in time and one spatial direction or alternatively static inhomogeneities varying in two spatial directions.

The case of a static one-dimensional magnetic field inhomogeneity has been discussed in detail in \cite{Gies:2013yxa},
where two different situations compatible with a global projector were identified.
Here we slightly generalize these situations to allow also for insights into higher dimensional and spatio-temporal field inhomogeneities.
These situations require special alignments of the direction of the inhomogeneity $\pmb{\nabla}{\cal E}(x)$ with respect to the magnetic field vector $\hat{\vec B}$:
\begin{gather}
 (i):\quad k_\parallel^\mu\partial_\mu{\cal E}(x)=0 \quad \to\quad  P_{\parallel}^{\mu\nu}(k) = P_{\parallel}^{\mu\nu}(k')\equiv\tilde P_\parallel^{\mu\nu}\quad \text{and}\quad k_\parallel^2=k_\parallel'^2\,, \nonumber\\
 (ii):\quad k_\perp^\mu\partial_\mu{\cal E}(x)=0 \quad \to\quad  P^{\mu\nu}_{\perp}(k) = P^{\mu\nu}_{\perp}(k')\equiv \tilde P^{\mu\nu}_{\perp}\quad \text{and}\quad k_\perp^2=k_\perp'^2\,, \label{eq:Bcases} 
\end{gather}
In case $(i)$ the parallel momentum components are not affected by the inhomogeneity and correspondingly the parallel polarization mode does not sense the inhomogeneity.
For $(ii)$ the perpendicular momentum components remain invariant such that the perpendicular polarization mode is conserved.
Obviously only case $(ii)$, being independent of $\omega$, is compatible with a magnetic field inhomogeneity that varies in both space and time.
Here we can study 
an inhomogeneity which varies at most in $1+1$ dimensions with $\pmb{\nabla}{\cal E}(x)\sim\hat{\vec B}$.
Assuming the spatial variation along ${\rm x}$ such that the profile of the inhomogeneity is given by ${\cal E}({\rm x},t)$ this implies that $\hat{\vec B}\sim\vec{e}_{\rm x}$.
Conversely, in case $(i)$ we can only study purely spatial -- at most two dimensional -- inhomogeneities. 
Aiming at an inhomogeneity which w.l.o.g. depends on the coordinates ${\rm x}$ and ${\rm y}$, i.e., ${\cal E}({\rm x},{\rm y})$, the condition $(i)$ in \Eqref{eq:Bcases} can only be fulfilled if $\hat{\vec B}\sim\vec{e}_{\rm z}$.

To keep the expressions as compact and transparent as possible, the
following explicit examples are limited to the leading contributions
in a perturbative expansion in powers of the amplitude of the
electromagnetic field inhomogeneity, i.e., to the
$\ell=1$ contribution in \Eqref{eq:SumPiinh}.  Such a truncation is also
well-justified from an experimental point of view, as the field
strengths available in high-intensity laser facilities all fulfill
$\frac{e{\cal E}}{m^2}\ll1$; cf. Sec.~\ref{seq:Exp} below.  All
relevant information about the photon polarization tensor in a
constant magnetic background field is recollected in
Appendix~\ref{app:B=const.}.

\begin{figure}
 \centering
  \includegraphics[width=0.42\textwidth]{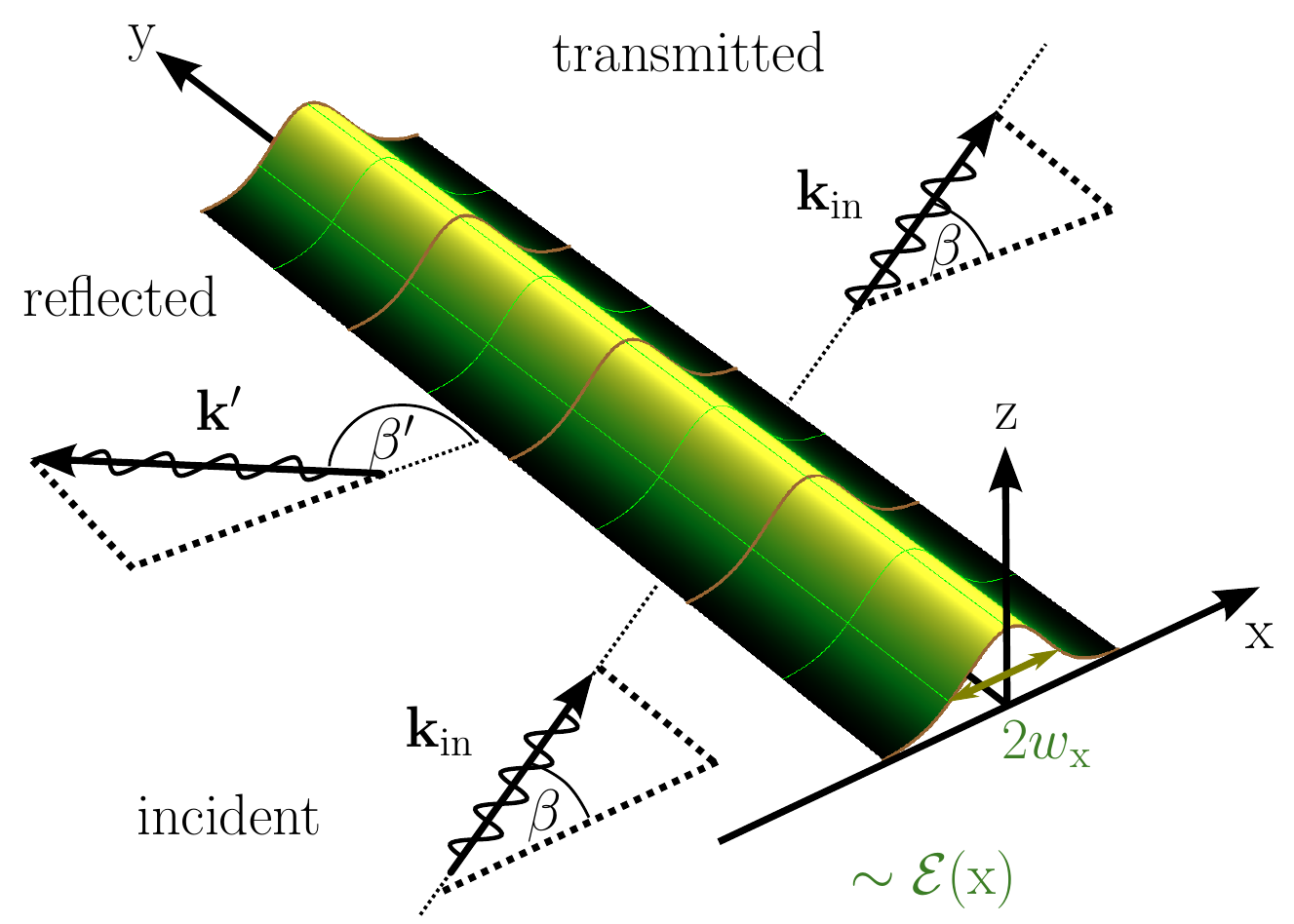} 
\caption{Schematic depiction of quantum reflection for a static one-dimensional field inhomogeneity \cite{Gies:2013yxa}.
The incident probe photons with four wave vector $k_{\rm in}^\mu=\omega_{\rm in}(1,\cos\beta,\sin\beta,0)$ propagate towards the inhomogeneity of amplitude profile ${\cal E}({{\rm x}})$ which asymptotically falls off to zero for large values of $|{{\rm x}}|$. The inhomogeneity is infinitely extended in the transversal directions.}
\label{fig:Qref}
\end{figure}

\subsubsection{Static one-dimensional inhomogeneity} \label{seq:1+0}

Let us briefly recall the results of \cite{Gies:2013yxa} for a static one-dimensional magnetic field inhomogeneity fulfilling $\partial_\mu{\cal E}(x) \sim \vec{e}_{\rm x}$, depicted in Fig.~\ref{fig:Qref}.
As detailed above, we have identified two specific situations where a global projector $(\tilde P_p)^{\mu\nu}$ can be identified.
In case $(i)$ we can give results for probe photons in the $p=\parallel$ polarization mode. Contrarily, in case $(ii)$ this is possible for $p=\perp$. 
For these two cases, the induced photon field~\eqref{eq:Aout_1d_Pi} is given by
\begin{multline}
 A^{\rm out}_p(x)=iA^{\rm in}_p(x)\,\frac{\alpha}{\pi}\,\frac{c_p}{90}\,\frac{\sin^2\varangle(\vec{B},\vec{k}_{\rm in})}{|\cos\beta|}\,\omega_{\rm in} \\
 \times\Biggl\{\int{\rm dx'}\left(\frac{e{\cal E}({\rm x}')}{m^2}\right)^{2}
 +{\rm e}^{-i(2\omega_{\rm in}\cos\beta){\rm x}}\int{\rm dx'}\,{\rm e}^{i(2\omega_{\rm in}\cos\beta){\rm x}'}\left(\frac{e{\cal E}({\rm x}')}{m^2}\right)^{2}\Biggr\} + {\cal O}\bigl((\tfrac{e\mathcal{E}}{m^2})^4\bigr) \,, \label{eq:Aout1d}
\end{multline}
with numerical coefficients $c_\parallel=7$, $c_\perp=4$. The corresponding reflection coefficients read [cf. \Eqref{eq:R1d}]
\begin{equation}
 R_p = \left|\frac{\alpha}{\pi}\,\frac{c_p}{90}\,\frac{\sin^2\varangle(\vec{B},\vec{k}_{\rm in})}{\cos\beta}\,\omega_{\rm in}\int{\rm d}{\rm x}'\,{\rm e}^{i(2\omega_{\rm in}\cos\beta'){\rm x}'} \left(\frac{e\cal{E}({\rm x}')}{m^2}\right)^{2}
\right|^2 + {\cal O}\bigl((\tfrac{e\cal{E}}{m^2})^{6}\bigr)\,. \label{eq:res}
\end{equation}
As discussed already in \cite{Gies:2014jia}, \Eqref{eq:res} diverges for $\beta\to\pm\frac{\pi}{2}$.
This unphysical behavior can be attributed to the unphysical limit of an infinitely long interaction of the probe photons and the inhomogeneity at ``grazing incidence'' (cf. also Sec.~\ref{seq:Exp}).
A perturbative treatment of the induced photon field, as employed here, of course can only be justified for $R_p \ll 1$.

\subsubsection{1+1 dimensional spatio-temporal inhomogeneity} \label{seq:1+1}

Apart from the static one-dimensional inhomogeneity considered in Sec.~\ref{seq:1+0}, case $(ii)$ is also compatible with a $1+1$ dimensional inhomogeneity
of amplitude profile ${\cal E}({\rm x},t)$.
For this background field,
Eqs.~\eqref{eq:Aindout} and \eqref{eq:Pip_speckin} result in
\begin{multline}
 A^{\rm out}_\perp(x)= ia(\omega_{\rm in}){\rm e}^{i\omega_{\rm in}\sin\beta y}\frac{2}{45}\frac{\alpha}{\pi} \, \omega_{\rm in}^2\sin^2\beta\,\int\frac{{\rm d}k_{\rm x}}{2\pi}\,\frac{1}{\omega(k_{\rm x})}\,{\rm e}^{i(k_{\rm x}{\rm x}-\omega(k_{\rm x})t)} \\
 \times \int{\rm d}{\rm x}'\int{\rm d}t'\,{\rm e}^{i[(\omega_{\rm in}\cos\beta-k_{\rm x}){\rm x}'-(\omega_{\rm in}-\omega(k_{\rm x}))t']} \left(\frac{e{\cal E}({\rm x}',t')}{m^2}\right)^{2} + {\cal O}\bigl((\tfrac{e{\cal E}}{m^2})^4\bigr)\,, \label{eq:Aout1+1}
\end{multline}
with $\omega(k_{\rm x})=\sqrt{k_{\rm x}^2+\omega_{\rm in}^2\sin^2\beta}$.

Equation \eqref{eq:Aout1+1} can readily be used for general field
inhomogeneities ${\cal E}({\rm x},t)$. For simplicity, we limit
ourselves to field inhomogeneities of the type $\mathcal{E}({\rm x},t)
= \mathcal{E}({\rm x}) \cos(\Omega t)$, with frequency $\Omega\geq0$.
This scenario resembles Fig.~\ref{fig:Qref}, with the inhomogeneity
featuring an additional global harmonic time dependence
$\sim\cos(\Omega t)$. The Fourier integral over time can be performed
straightforwardly, resulting in
\begin{equation}
 \int\!{\rm d}t'\,{\rm e}^{-i(\omega_{\rm in}-\omega(k_{\rm x})) t'} \cos^2(\Omega t')
  =\frac{\pi}{2}\!\sum_{n=-1}^{+1} (1+\delta_{n0})\,\Theta\bigl(\omega_{\rm in} (1-|\sin\beta|) - 2n\Omega \bigr)\frac{\omega(k_{\rm x})}{|k_{{\rm x},2n}|}\Bigl[\delta\bigl(k_{\rm x}-|k_{{\rm x},2n}|\bigr)+\delta\bigl(k_{\rm x}+|k_{{\rm x},2n}|\bigr)\Bigr] , \label{eq:id1}
\end{equation}
where we have rewritten the $\delta$ functions for the energy in terms of $\delta$ functions for $k_{\rm x}$. Here $\delta_{nn'}$ is the Kronecker delta
and $k_{{\rm x},2n}^2\equiv(\omega_{\rm in}-2n\Omega)^2-\omega_{\rm in}^2\sin^2\beta$. 
Correspondingly, the integrals in \Eqref{eq:Aout1+1} can be reduced to a single one.
Upon insertion of \Eqref{eq:id1} into \Eqref{eq:Aout1+1}, we obtain
\begin{multline}
 A^{\rm out}_\perp(x)= ia(\omega_{\rm in}){\rm e}^{i\omega_{\rm in}{\rm y}\sin\beta}\frac{1}{90}\frac{\alpha}{\pi} \, \omega_{\rm in}^2\sin^2\beta \sum_{n=-1}^{+1} (1+\delta_{n0})\,
 \Theta\bigl(\omega_{\rm in}(1 - |\sin\beta|) - 2n\Omega\bigr)\,\frac{1}{|k_{{\rm x},2n}|}\,{\rm e}^{i|k_{{\rm x},2n}|{\rm x}-i(\omega_{\rm in}-2n\Omega)t} \\
 \times 
 \Bigg\{\int{\rm d}{\rm x}'\,{\rm e}^{i(\omega_{\rm in}\cos\beta-|k_{{\rm x},2n}|){\rm x}'} \left(\frac{e{\cal E}({\rm x}')}{m^2}\right)^2
 +{\rm e}^{-2i|k_{{\rm x},2n}|{\rm x}}\int{\rm d}{\rm x}'\,{\rm e}^{i(\omega_{\rm in}\cos\beta+|k_{{\rm x},2n}|){\rm x}'} \left(\frac{e{\cal E}({\rm x}')}{m^2}\right)^2\Biggr\}
   + {\cal O}\bigl((\tfrac{e{\cal E}}{m^2})^4\bigr)\,, \label{eq:Aout1+1magnetic}
\end{multline}
which is still applicable for arbitrarily shaped amplitude profiles ${\cal E}({\rm x})$ compatible with the locally constant field approximation.

For a transparent discussion (cf. Fig.~\ref{fig:2dim}), it is convenient to distinguish between induced photon fields propagating in forward and backward direction, respectively:
We consider a given induced wave component as propagating in forward direction if the ${\rm x}$ component of its wave vector $k_{\rm x}'=\pm|k_{{\rm x},2n}|$ fulfills ${\rm sign}(k'_{\rm x}k_{\rm in,x})={\rm sign}(k'_{\rm x}\cos\beta)>0$, and as propagating in backward direction if ${\rm sign}(k'_{\rm x}\cos\beta)<0$.
Clearly, the induced photon field~\eqref{eq:Aout1+1magnetic} is made up of various wave components propagating in forward and backward direction.

As expected from its harmonic oscillation in time with frequency $\Omega$, the background field can absorb and supply energy to the probe photons.
Moreover, as $\cos^2\chi=\frac{1}{2}+\frac{1}{4}({\rm e}^{2i\chi}+{\rm e}^{-2i\chi})$, we always encounter elastic contributions resembling \Eqref{eq:Aout1d} without energy transfer from the inhomogeneity.
Correspondingly, the induced photon field is a superposition of plane waves with different positive frequencies $\omega'_{2n}=\omega_{\rm in}-2n\Omega$
and propagation directions $\vec{k}'_{\pm, 2n}=(\pm|k_{{\rm x},2n}|,\omega_{\rm in}\sin\beta,0)$, where $n\in\{0,\pm1\}$.
As these waves propagate on the light cone and the momentum component along $\vec{e}_{\rm y}$ is conserved, a change in frequency inevitably alters the $\rm x$ component of their wave vector and thus their propagation direction.
\begin{figure}
 \centering
  \includegraphics[width=0.48\textwidth]{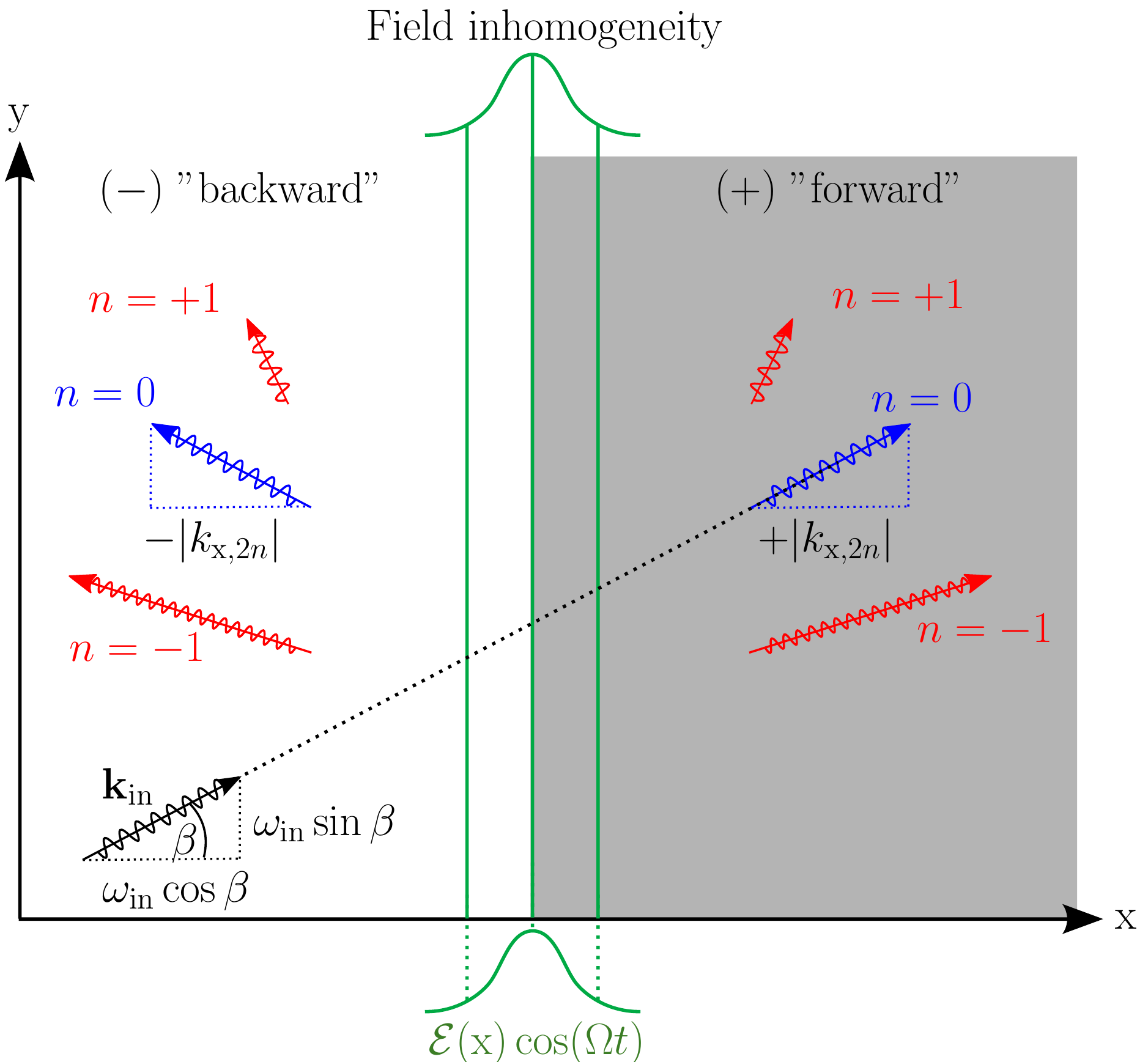} 
\caption{Quantum reflection for a $1+1$ dimensional field
  inhomogeneity featuring a harmonic time dependence $\sim\cos(\Omega
  t)$.  The basic scenario is the same as depicted in
  Fig.~\ref{fig:Qref} and the inhomogeneity is infinitely extended in
  the transversal directions.  However, due to the harmonic
  oscillation with $\Omega$, apart from the elastic contributions
  ($n=0$), also inelastic contributions ($n=\pm1$) with frequencies
  $\omega'_{2n}=\omega_{\rm in}-2n\Omega$ and propagation directions
  $\vec{k}'_{\pm, 2n}=(\pm|k_{{\rm x},2n}|,\omega_{\rm
    in}\sin\beta,0)$ are induced (cf. main text).  We label induced
  photon waves propagating into the half-space right of the
  inhomogeneity with $(+)$ and those propagating into the left
  half-space with $(-)$.  For convenience we will also speak of
  ``backward'' (white) and ``forward'' (gray-shaded) directions: If
  the ${\rm x}$ momentum component of a given induced contribution has
  the opposite (same) sign as the ${\rm x}$ momentum component of the
  incident probe photon, it is induced in backward (forward)
  direction.}
\label{fig:2Dplane}
\end{figure}
However, in the limit $\Omega\to0$ we recover \Eqref{eq:Aout1d} with $p=\perp$, but $\sin^2\varangle(\vec{B},\vec{k}_{\rm in})\to\sin^2\beta$, with the latter substitution accounting for the different kinematic situation considered here;
the three photon components induced in forward and backward direction then merge to form a single transmitted and reflected photon wave, respectively.

In close analogy to \Eqref{eq:res} above,
for a given induced photon wave of frequency $\omega'_{2n}$ we define reflection -- or more generally scattering -- coefficients $R^{(\pm)}_{2n}$ as the modulus squared of the respective induced photon field normalized by $a(\omega_{\rm in})$.
For \Eqref{eq:Aout1+1magnetic} this results in
\begin{equation}
 R_{\perp,2n}^{(\pm)}=\Theta\bigl(\omega_{\rm in}(1-|\sin\beta|) - 2n\Omega \bigr)\left|\frac{\alpha}{\pi} \frac{1+\delta_{n0}}{90} \,\frac{\omega_{\rm in}^2\sin^2\beta}{k_{{\rm x},2n}} 
\int{\rm d}{\rm x}'\,{\rm e}^{i(\omega_{\rm in}\cos\beta\pm|k_{{\rm x},2n}|){\rm x}'} \left(\frac{e{\cal E}({\rm x}')}{m^2}\right)^2\right|^2
   + {\cal O}\bigl((\tfrac{e{\cal E}}{m^2})^6\bigr)\,. \label{eq:1+1res}
\end{equation}

The Heaviside function only allows for contributions satisfying $\omega'_{2n}\geq\omega_{\rm in}|\sin\beta|$.
For $n\in\{-1,0\}$ this condition is trivially fulfilled. 
By contrast, this condition becomes nontrivial for $n=1$:
incident probe photons of energy $\omega_{\rm in}$ can induce photons with $n=1$ only if the modulus of the photon energy after emission of two quanta of energy $\Omega$ is still large enough to facilitate induced outgoing photons propagating on the light cone while simultaneously ensuring momentum conservation along ${\rm y}$.

\subsubsection{Static two-dimensional inhomogeneity} \label{seq:2+0}

\begin{figure}[htpb]
 \centering
  \includegraphics[width=0.36\textwidth]{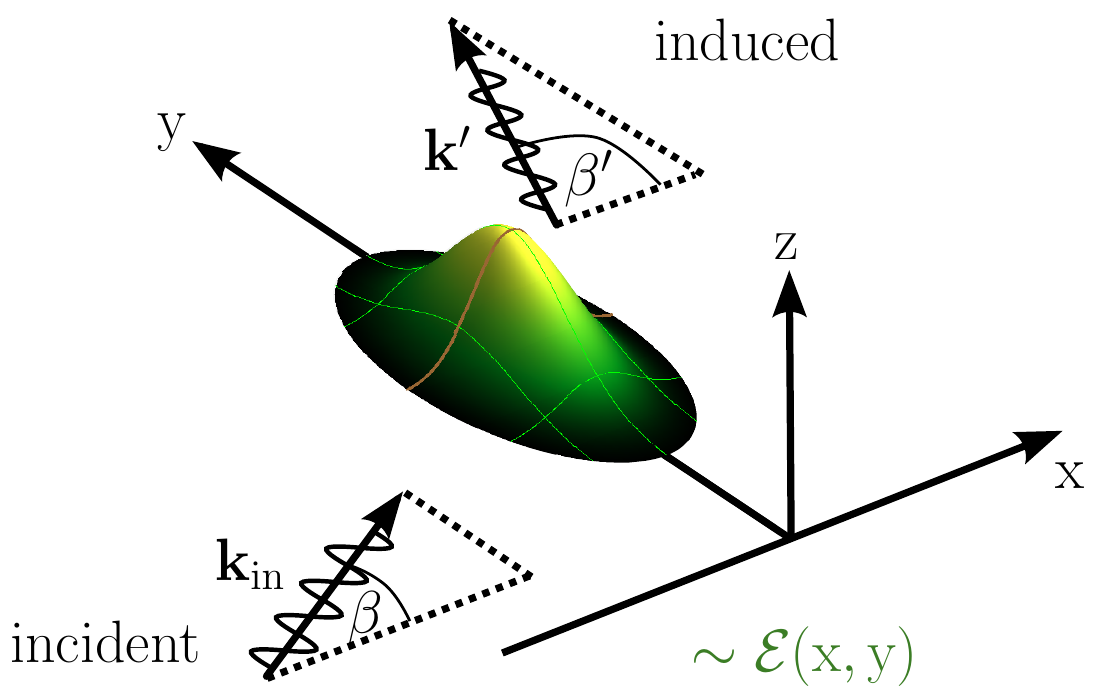} 
\caption{Visualization of quantum reflection in the case of a localized static two-dimensional field inhomogeneity. Upon interaction with the inhomogeneity, the incident probe photons with wave vector $\vec{k}_{\rm in}=\omega_{\rm in}(\cos\beta,\sin\beta,0)$ induce outgoing circular photon waves with wave vector $\vec{k}'=\omega_{\rm in}(\cos\beta',\sin\beta',0)$. 
As the inhomogeneity is independent of time, the energy of the induced photons equals that of the incident probe photons.}
\label{fig:2dim}
\end{figure}

Case $(i)$ also allows for localized static two-dimensional inhomogeneities of amplitude profile ${\cal E}({\rm x},{\rm y})$ (cf. Fig.~\ref{fig:2dim}).
Inserting this profile into Eqs.~\eqref{eq:Aindout} and \eqref{eq:Pip_speckin} and performing the integrations over $k_{\rm x}$ and $k_{\rm y}$ by resorting to polar coordinates,
we obtain the following compact expression for the induced photon field,
\begin{equation}
 A^{\rm out}_\parallel(x)
   =i a(\omega_{\rm in})\,{\rm e}^{-i\omega_{\rm in}t}\frac{7}{90} \frac{\alpha}{\pi} \, \omega_{\rm in}^2
 \int{\rm d}^2{\vec x}'\, {\rm e}^{i\omega_{\rm in}[\cos\beta{\rm x}'+\sin\beta{\rm y}']}\, J_0(\omega_{\rm in}|\vec{x}-\vec{x}'|)
 \left(\frac{e{\cal E}({\rm x}',{\rm y}')}{m^2}\right)^{2}  + {\cal O}\bigl((\tfrac{e{\cal E}}{m^2})^4\bigr), \label{eq:Aout2d}
\end{equation}
with $\vec{x}=({\rm x},{\rm y},0)$,  $\vec{x}'=({\rm x}',{\rm y}',0)$ and $ \int{\rm d}^2{\vec x}'=\int{\rm d}{\rm x}'\int{\rm d}{\rm y}'$.
Here $J_n(\chi)$ is the Bessel function of the first kind of order $n\in\mathbb{N}_0$,
which for large real valued arguments $\chi$ behaves as (cf. formula 8.451.1 of \cite{Gradshteyn})
\begin{equation}
 J_n(\chi) =\frac{1}{i^{n+1}}\sqrt{\frac{i}{2\pi\chi}}\,{\rm e}^{i\chi}\,\bigl[1+i\,(-1)^n\,{\rm e}^{-2i\chi}\bigr]+{\cal O}\bigl(\chi^{-3/2}\bigr) . \label{eq:J0assympt}
\end{equation}

For a localized inhomogeneity, the integral in \Eqref{eq:Aout2d} only receives contributions from a limited range of $\vec{x}'$.
Conversely, detection is assumed to take place in the far field, i.e., far outside the regime where the integral over $\vec{x}'$ receives any substantial contributions (cf. Sec.~\ref{seq:Gen}).
Hence, the modulus $|\vec{x}-\vec{x}'|$ can be expanded as follows
$|\vec{x}-\vec{x}'|=|\vec{x}|\bigl(1-\frac{\vec{x}\cdot\vec{x}'}{|\vec{x}|^2}+{\cal O}(\varepsilon^2)\bigr)$, with $\varepsilon\equiv|\vec{x}'|/|\vec{x}|\ll1$.

Defining $\vec{k}'=\omega_{\rm in}\frac{\vec{x}}{|\vec{x}|}=\omega_{\rm in}(\cos\beta',\sin\beta',0)$
and adopting the expansion~\eqref{eq:J0assympt} in \Eqref{eq:Aout2d},
we approximate the factor $|\vec{x}-\vec{x}'|$ in the phase of the exponential as $|\vec{x}-\vec{x}'|\approx|\vec{x}|-\frac{1}{\omega_{\rm in}}\vec{k}'\cdot\vec{x}'$,
and that in the prefactor just by $|\vec{x}-\vec{x}'|\approx|\vec{x}|$.
Focusing on the contribution describing an outgoing circular wave only, we obtain
\begin{equation}
 A^{\rm out}_\parallel(x)
   \approx a(\omega_{\rm in})\,\frac{{\rm e}^{i\omega_{\rm in}(|\vec{x}|-t)}}{\sqrt{|\vec{x}|}}\,f_\parallel(\vec{k}',\vec{k}_{\rm in}) \,, \label{eq:aout}
\end{equation}
with scattering amplitude
\begin{equation}
 f_\parallel(\vec{k}',\vec{k}_{\rm in}) = \sqrt{\frac{i}{2\pi\omega_{\rm in}}}\,\frac{7}{90} \frac{\alpha}{\pi}\, \omega_{\rm in}^2 \int {\rm d}^2\vec{x}' \ e^{i (\vec{k}_{\rm in} - \vec{k}')\cdot\vec{x}' } \left( \frac{e\mathcal{E} ({\rm x}',{\rm y}')}{m^2} \right)^2 + {\cal O}\bigl((\tfrac{e{\cal E}}{m^2})^4\bigr). \label{eq:f(k,k)}
\end{equation}

For large far-field distances $|\vec{x}|$ about the scattering center, the modulus-squared of the induced photon field amplitude through a polar angle interval ${\rm d}\beta'$ is independent of $|\vec{x}|$, i.e.,
$|A^{\rm out}(x)|^2|\vec{x}|{\rm d}\beta'=|A^{\rm in}(x)|^2|f(\vec{k}',\vec{k}_{\rm in})|^2{\rm d}\beta'$.
Hence, the differential cross section for inducing outgoing photons propagating in direction $\vec{k}'$ is given by
 \begin{equation}
 \frac{{\rm d} \sigma }{{\rm d} \beta'}(\vec{k}',\vec{k}_{\rm in}) \equiv \frac{|A^{\rm out}(x)|^2|\vec{x}|}{|A^{\rm in}(x)|^2} = | f(\vec{k}',\vec{k}_{\rm in}) |^2 \, .	\label{eq:diffcross}
\end{equation}
The cross section for photons hitting a detector located at $\beta'$ and spanning an opening angle $\delta \beta'(\beta')$ about a given value of $\beta'$ is obtained straightforwardly by integration, $\sigma[\beta,\delta\beta'(\beta')] = \int_{\delta\beta'}{\rm d}\beta'\, (\frac{{\rm d}\sigma}{{\rm d}\beta'} )$.
Provided that the extent of the inhomogeneity is larger than the probe beam diameter $w_{\rm probe}$,
the differential number of induced outgoing photons per angle is given by
$\frac{{\rm d}N_{\rm out}}{{\rm d}\beta'}=\frac{N_{\rm probe}}{w_{\rm probe}}\frac{{\rm d}\sigma}{{\rm d}\beta'}$,
and the number of induced outgoing photons corresponding to a cross section $\sigma[\beta,\delta\beta'(\beta')]$ by $N_{\rm out}[\beta,\delta\beta'(\beta')]=\frac{N_{\rm probe}}{w_{\rm probe}}\,\sigma[\beta,\delta\beta'(\beta')]$.

For completeness, note that Eqs.~\eqref{eq:aout} and \eqref{eq:f(k,k)} resemble the result of a quantum mechanical calculation in the Born-approximation (cf. also the quantum mechanical analogy presented in \cite{Gies:2013yxa}).
We can easily vary detector sizes and analyze the fraction of scattered photons for all angles $-\pi<\beta'\leq\pi$.

In the standard language of scattering theory, the angle $\beta'$ is related to the scattering angle $\theta$ by $\theta=\beta'-\beta$, where $\theta$ denotes the angle between $\vec{k}_{\rm in}$ and $\vec{k}'$.
As the process is elastic, we have $|\vec{k}_{\rm in}|=|\vec{k}'|$, such that the scattering amplitude is in fact only a function of $\omega_{\rm in}$, $\theta$ and, of course, the parameters of the inhomogeneity (including $\beta$).

\subsection{Crossed fields} \label{seq:Cross}

We now focus on crossed fields, i.e., a class of electromagnetic field inhomogeneities featuring both electric and magnetic field components. The electric and magnetic fields are assumed to be orthogonal to each other, i.e., $\hat{\vec{E}}\perp\hat{\vec{B}}$ with fixed unit vectors $\hat{\vec{E}}$ and $\hat{\vec{B}}$, having the same amplitude profile, i.e., $\vec{E}(x)={\cal E}(x)\hat{\vec{E}}$ and $\vec{B}(x)={\cal E}(x)\hat{\vec{B}}$.
This scenario is, e.g., applicable to linearly polarized plane wave fields.

As outlined above, an inhomogeneous field configuration can only affect those momentum components which sense the inhomogeneity.
Thus, a convenient strategy to find invariant projection operators $\tilde P^{\mu\nu}$ for a given inhomogeneity of profile $\cal E$ is to identify specific kinematic situations  
for which at least one projector is insensitive to changes in the momentum components associated with the coordinates of the inhomogeneity.

On the light cone, the photon polarization tensor in constant crossed fields is again spanned by just two projectors denoted by $P_p^{\mu\nu}(k)$ with $p\in\{1,2\}$ [cf. \Eqref{eq:P_12}], projecting onto photon polarization modes.
As the explicit expressions for these projectors are less intuitive as in the magnetic field case, we relegate them to App.~\ref{app:conscrossfields}, where also all necessary details for the polarization tensor in constant crossed fields are recollected. 
For probe photons polarized in mode $p=1$, the electric field component of the photon field is perpendicular to $\hat{\vec{E}}$, and lies in the plane spanned by the photon's wave vector $\vec{k}$ and $\hat{\vec{B}}$. For mode $p=2$, the roles of the electric and magnetic background field components are interchanged \cite{Dittrich:2000zu}.

\subsubsection{Static one-dimensional inhomogeneity} \label{seq:c1+0}

We find that $P_1^{\mu\nu}(k)$ is invariant under $\vec{k}\to\vec{k} + c(k) \hat{\vec E}$, with $c(k)$ denoting an arbitrary function of the components of $k^\mu$.
Analogously $P_2^{\mu\nu}(k)$ is invariant under $\vec{k} \to \vec{k} + c(k) \hat{\vec B}$.
Hence, for $\pmb{\nabla}{\cal E}\sim\hat{\vec E}$ the projector $P_1^{\mu\nu}$ corresponds to a global projector, while for $\pmb{\nabla}{\cal E}\sim\hat{\vec B}$ the projector $P_2^{\mu\nu}$ is a global projector. 

For a one-dimensional spatial field inhomogeneity ${\cal E}({\rm x})$,
the corresponding reflection coefficients essentially agree with \Eqref{eq:res}.
They are obtained by the substitution $\sin^2\varangle(\vec{B},\vec{k}_{\rm in})\to (1 - \vec{s}\cdot\vec{e}_{\rm y}\sin\beta)^2$, with $\vec{s}=\hat{\vec E}\times\hat{\vec B}$, accounting for the different kinematic situation. Also, we then have $c_1=7$, $c_2=4$. 
Again, the reflection coefficients generically diverge for $\beta \rightarrow \pm \frac{\pi}{2}$.
However, for the special case of $\vec{s}={\vec e}_{\rm y}$ only one divergent direction remains, while the reflection coefficient approaches zero for $\beta \rightarrow \frac{\pi}{2}$.
The reason for this is that the propagation of light is not affected by weak crossed background fields, if $( \hat{\vec k},\hat{\vec E},\hat{\vec B})$ form the basis of an ortho-normalized coordinate system \cite{Dittrich:2000zu}.

We skip the discussion of a $1+1$ dimensional inhomogeneity for crossed fields which would follow  Sec.~\ref{seq:1+1} and directly jump to a $2+1$ dimensional inhomogeneity.

\subsubsection{2+1 dimensional spatio-temporal inhomogeneity} \label{seq:c2+1}

In sharp contrast to the case of a purely magnetic field, the projectors can be rendered momentum independent for crossed fields by setting a single spatial momentum component to zero, say $k_{\rm z}=0$,
and adjusting the probe photon polarization accordingly.

Without loss of generality we assume $\vec{s}\equiv\hat{\vec E}\times\hat{\vec B}=\vec{e}_{\rm y}$, and for the remainder of this paper limit ourselves to the two special cases
\begin{equation}
  \{p=1\,,\ \hat{\vec E}=-\vec{e}_{\rm x} \,,\ \hat{\vec B}=+\vec{e}_{\rm z}\} \quad\text{and}\quad \{p=2\,,\  \hat{\vec E}=+\vec{e}_{\rm z} \,,\ \hat{\vec B}=+\vec{e}_{\rm x}\}\,.
\label{eq:spec}
\end{equation}
These are chosen such that the associated projectors are energy-independent and independent of the kinematics in the ${\rm x}$-${\rm y}$ plane (cf. Appendix~\ref{app:conscrossfields}). Correspondingly, in the first (second) case $P_1^{\mu\nu}$ ($P_2^{\mu\nu}$) amounts to a global projector,
implying that in these cases we can consider spatio-temporal field inhomogeneities of amplitude profile ${\cal E}({\rm x},{\rm y},t)$.
To keep notations compact, we label our results by $p$ only. One should, however, keep in mind the additional constraints on the field orientations~\eqref{eq:spec}.
For a field inhomogeneity of this type, Eqs.~\eqref{eq:Aindout} and \eqref{eq:Pip_speckin} result in
\begin{multline}
 A^{\rm out}_p(x)=i a(\omega_{\rm in})\frac{c_p}{90} \frac{\alpha}{\pi} \, \int\frac{{\rm d}k_{\rm x}}{2\pi}\int\frac{{\rm d}k_{\rm y}}{2\pi}\,{\rm e}^{i[k_{\rm x}{\rm x}+k_{\rm y}{\rm y}-\sqrt{k_{\rm x}^2+k_{\rm y}^2}t]}\,\frac{1}{2\sqrt{k_{\rm x}^2+k_{\rm y}^2}}\,\Bigl[\omega_{\rm in}^2(1-\sin\beta)^2+\bigl(\sqrt{k_{\rm x}^2+k_{\rm y}^2}-k_{\rm y}\bigr)^2\Bigr] \\
 \times\int {\rm d}t'\int{\rm d}{\rm x}'\int{\rm d}{\rm y}'\,{\rm e}^{i[(\omega_{\rm in}\cos\beta-k_{\rm x}){\rm x}'+(\omega_{\rm in}\sin\beta-k_{\rm y}){\rm y}'-(\omega_{\rm in}-\sqrt{k_{\rm x}^2+k_{\rm y}^2})t']}\left(\frac{e{\cal E}({\rm x}',{\rm y}',t')}{m^2}\right)^{2}+{\cal O}\bigl((\tfrac{e{\cal E}}{m^2})^4\bigr)\,, \label{eq:Aoutgencrossed1+2}
\end{multline}
where $c_1=7$, $c_2=4$ are the numerical coefficients associated with the two cases in \Eqref{eq:spec}.

\begin{figure}[h]
\center
\includegraphics[width=0.45\textwidth]{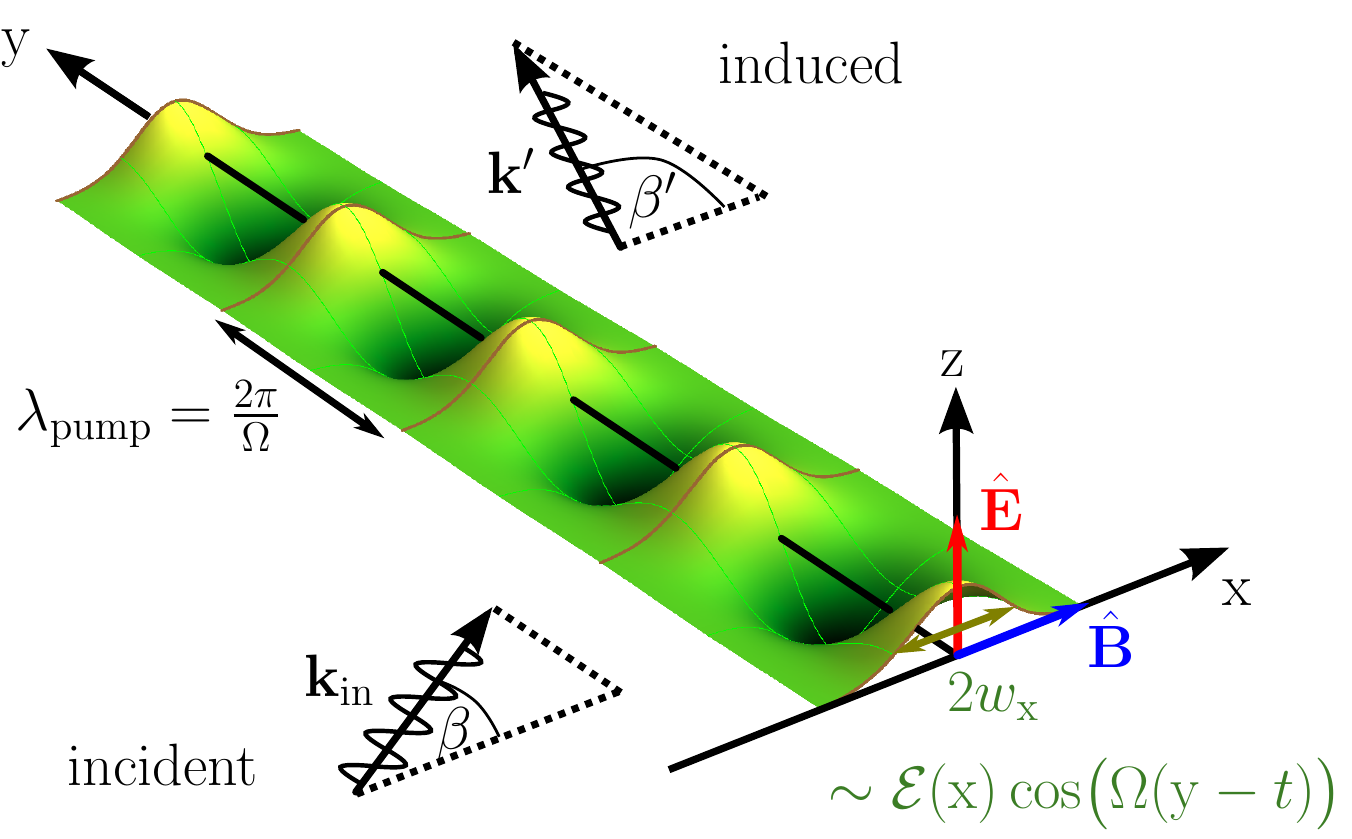}
\caption{Schematic depiction of quantum reflection for a $2+1$ dimensional field inhomogeneity ${\cal E}({\rm x},{\rm y},t)={\cal E}({\rm x})\cos\bigl(\Omega({\rm y}-t)\bigr)$ for $p=1$.
This situation grasps the basic features of the experimental scenario of quantum reflection in a high-intensity laser set-up.
In the vicinity of its beam waist the pump field configuration is well compatible with such a field profile (cf. Sec.~\ref{seq:Exp}).
Incident probe photons of four momentum $k_{\rm in}^\mu$ traveling in the $\rm x$-$\rm y$ plane induce outgoing photons of four momentum $k'^\mu$.
As the spacetime dependence of the field inhomogeneity allows for both energy and momentum exchange between the probe and pump fields,
energy and momentum, and thus also direction of the induced photons in general differ from that of the incident probe photons.}
\label{fig:laser}
\end{figure}

A much more handy expression is obtained when the amplitude profile of the field inhomogeneity factorizes into a {\it longitudinal} ``plane wave'' profile $\sim\cos(\Omega({\rm y}-t))$ and a {\it transversal} profile ${\cal E}({\rm x})$.
This is assumed in the following, ${\cal E}({\rm x},{\rm y},t)={\cal
  E}({\rm x})\cos(\Omega({\rm y}-t))$, see also Fig.~\ref{fig:laser}.
This has several advantages: First, from a practical point of view,
the five integrals to be performed in \Eqref{eq:Aoutgencrossed1+2} can
be reduced to a single remaining integral in this case. The
integrations over ${\rm y}'$ and $t'$ yield $\delta$ functions which
in turn render the integrations over $k_{\rm x}$ and $k_{\rm y}$
trivial.  Second, this amplitude profile actually approximates the electromagnetic field inhomogeneities
generated by high-intensity lasers (cf. also Sec.~\ref{seq:Exp}
below).

With the help of the following identity, somewhat reminiscent of \Eqref{eq:id1} above,
\begin{multline}
 \int{\rm d}t\int{\rm dy}\,{\rm e}^{i[(\omega_{\rm in}\sin\beta-k_{\rm y}){\rm y}-(\omega_{\rm in}-\sqrt{k_{\rm x}^2+k_{\rm y}^2})t]}\cos^2\bigl(\Omega({\rm y}-t)\bigr) \\
 =\pi^2\sum_{n=-1}^{+1} (1+\delta_{n0})\,\Theta\bigl(k_{{\rm x},2n}^2\bigr)\,\frac{\sqrt{k_{\rm x}^2+k_{\rm y}^2}}{|k_{{\rm x},2n}|}\,\delta\bigl(k_{\rm y}-\omega_{\rm in}\sin\beta+2n\Omega\bigr)\,\Bigl[\delta\bigl(k_{\rm x}-|k_{{\rm x},2n}|\bigr)+\delta\bigl(k_{\rm x}+|k_{{\rm x},2n}|\bigr)\Bigr] ,
\end{multline}
where now $k_{{\rm x},2n}^2\equiv(\omega_{\rm in}-2n\Omega)^2-(\omega_{\rm in}\sin\beta-2n\Omega)^2$, we obtain
\begin{multline}
 A^{\rm out}_p(x)
 =i a(\omega_{\rm in})\frac{c_p}{360} \frac{\alpha}{\pi} \, \omega_{\rm in}^2(1-\sin\beta)^2
 \sum_{n=-1}^{+1} (1+\delta_{n0})\,\Theta\bigl(k_{{\rm x},2n}^2\bigr) \frac{1}{|k_{{\rm x},2n}|}\,{\rm e}^{i[|k_{{\rm x},2n}|{\rm x}+(\omega_{\rm in}\sin\beta-2n\Omega){\rm y}-(\omega_{\rm in}-2n\Omega)t]} \\
 \times\Biggl\{\int{\rm d}{\rm x}'{\rm e}^{i(\omega_{\rm in}\cos\beta-|k_{{\rm x},2n}|){\rm x}'}\left(\frac{e{\cal E}({\rm x}')}{m^2}\right)^{2}
 +{\rm e}^{-2i|k_{{\rm x},2n}|{\rm x}}\int{\rm d}{\rm x}'{\rm e}^{i(\omega_{\rm in}\cos\beta+|k_{{\rm x},2n}|){\rm x}'}\left(\frac{e{\cal E}({\rm x}')}{m^2}\right)^{2}\Biggr\}+{\cal O}\bigl((\tfrac{e{\cal E}}{m^2})^4\bigr)\,, \label{eq:A2+1res}
\end{multline}
made up of induced photon waves of positive frequencies $\omega'_{2n}=\omega_{\rm in}-2n\Omega$ and wave vectors $\vec{k}'_{\pm, 2n}=(\pm|k_{{\rm x},2n}|,\omega_{\rm in}\sin\beta-2n\Omega,0)$, with $n\in\{0,\pm1\}$.
The basic scenario resembles Fig.~\ref{fig:2dim} above.
The corresponding scattering coefficients are given by (cf. Sec.~\ref{seq:1+1})
\begin{equation}
 R_{p,2n}^{(\pm)}=\Theta\bigl(k_{{\rm x},2n}^2\bigr)\left|\frac{\alpha}{\pi}\frac{c_p(1+\delta_{n0})}{360}  
 \,\frac{\omega_{\rm in}^2(1-\sin\beta)^2}{k_{{\rm x},2n}}\int{\rm d}{\rm x}'{\rm e}^{i(\omega_{\rm in}\cos\beta\pm|k_{{\rm x},2n}|){\rm x}'}\left(\frac{e{\cal E}({\rm x}')}{m^2}\right)^{2}\right|^2+{\cal O}\bigl((\tfrac{e{\cal E}}{m^2})^6\bigr)\,. \label{eq:2+1res}
\end{equation}
Taking into account that $\omega_{\rm in}>0$ and $\Omega\geq0$,
the restriction to $k_{{\rm x},2n}^2 \geq 0$ implemented by the Heaviside function can alternatively be expressed as $\frac{2n\Omega}{\omega_{\rm in}}\leq\frac{1+\sin\beta}{2}$.
For $n\in\{-1,0\}$ this condition is always fulfilled, as then $\frac{2n\Omega}{\omega_{\rm in}}<0$, while $0<\frac{1+\sin\beta}{2}<1$.
Conversely, for $n=1$ the left-hand side of the above condition becomes positive $\frac{2\Omega}{\omega_{\rm in}}\geq0$.
This implies that we have to ensure that $\frac{2\Omega}{\omega_{\rm in}}\leq\frac{1+\sin\beta}{2}$.
For $\frac{2\Omega}{\omega_{\rm in}}\geq1$ this inequality can no longer be fulfilled, irrespectively of the value of $\beta$ in order to facilitate any induced photon waves with $n=1$ for a given angle of incidence $\beta$.
The physical reason for this is that under these circumstances the probe photon energy is too small to allow for an energy transfer of $2\Omega$ to the background field.

For completeness, we remark that the results derived in this section are also attainable from the photon polarization tensor in a linearly polarized plane wave field,
which is a special case of the the polarization tensor in a generic, elliptically polarized monochromatic plane wave background \cite{Baier:1975ff,Becker:1974en}.
A particularly convenient representation tailored to soft photon fields and perturbative weak field strengths is given in \cite{Gies:2014jia}.
The only difference to the procedure outlined in Sec.~\ref{seq:Gen} to build in the field inhomogeneity in the constant field polarization tensor {\it a posteriori}, 
is that the polarization tensor in a linearly polarized plane wave already accounts for the longitudinal profile of the field inhomogeneity $\sim\cos(\Omega({\rm y}-t))$.
Hence, only the transversal profile ${\cal E}({\rm x})$ has to be built in along the lines outlined in Sec.~\ref{seq:Gen}.
At ${\cal O}((e{\cal E})^2)$ and on the light cone (in the notation of \cite{Gies:2014jia}: $k_1^2=k_2^2=0$), the tensor structure of the photon polarization tensor in a linearly polarized plane wave (i.e., either $\xi_1$ or $\xi_2$ of \cite{Gies:2014jia} vanishes) is spanned by just two projection operators, conventionally represented as $\Lambda_1^\mu\Lambda_1^\nu$ and $\Lambda_2^\mu\Lambda_2^\nu$, with normalized four vectors $\Lambda_1^2=\Lambda_2^2=1$ \cite{Gies:2014jia}.
In fact these projectors can be identified with the projectors $P_1^{\mu\nu}$ and $P_2^{\mu\nu}$ for constant crossed fields.

\subsubsection{Static two-dimensional inhomogeneity} \label{seq:c2+0}

In close analogy to our considerations for a static two-dimensional, purely magnetic field inhomogeneity in Sec.~\ref{seq:c2+0},
we can also investigate the case of a two-dimensional inhomogeneity ${\cal E}({\rm x},{\rm y})$ featuring crossed fields.
Specializing to $\vec{s}=\vec{e}_{\rm y}$ and the special cases in \Eqref{eq:spec}
we obtain (cf. Sec.~\ref{seq:2+0} and Fig.~\ref{fig:2dim})
\begin{equation}
 A^{\rm out}_p(x)
   \approx a(\omega_{\rm in})\,\frac{{\rm e}^{i\omega_{\rm in}(|\vec{x}|-t)}}{\sqrt{|\vec{x}|}}\,f_p(\vec{k}',\vec{k}_{\rm in}) \,, \label{eq:aout_2}
\end{equation}
with scattering amplitude
\begin{equation}
 f_p(\vec{k}',\vec{k}_{\rm in}) = \sqrt{\frac{i}{2\pi\omega_{\rm in}}}\,\frac{c_p}{90} \frac{\alpha}{\pi}\,\frac{(1-\sin\beta)^2+(1 -\sin\beta')^2}{2}\, \omega_{\rm in}^2 \int {\rm d}^2\vec{x}' \ e^{i (\vec{k}_{\rm in} - \vec{k}')\cdot\vec{x}' } \left( \frac{e\mathcal{E} ({\rm x}',{\rm y}')}{m^2} \right)^2 + {\cal O}\bigl((\tfrac{e{\cal E}}{m^2})^4\bigr)\,. \label{eq:f(k,k)_2}
\end{equation}

\section{Towards experimental estimates} \label{seq:Exp}

Let us now discuss the results more quantitatively, adopting real
laser parameters.  We consider an experimental scenario of an
all-optics pump-probe type experiment employing two high intensity
lasers.  An electromagnetic field inhomogeneity is generated in the
focal spot of the {\it pump} laser, and probed by a second {\it
  probe} laser.  The probe beam is assumed to intersect the pump laser
beam right at its waist under an angle of $\beta$
(cf. Fig.~\ref{fig:laser}).  The number of induced photons is measured
by a detector in the field free region, thereby facilitating a clear
geometric signal to background separation.

The electromagnetic field pulses as generated by high intensity lasers evolve along the well-defined envelope of a Gaussian beam, whose transversal profile is of Gaussian type.
Correspondingly, we exclusively limit ourselves to transversal field profiles ${\cal E}({\rm x})={\cal E}\,{\rm e}^{-(2{\rm x}/w_{\rm x})^2}$ of amplitude $\cal E$, with $w_{\rm x}$ denoting the full width at $1/{\rm e}$ of its maximum.
Of course, an optimization of amplitude profiles could certainly allow for a substantial enhancement of the induced photon signal.
As a freely propagating Gaussian beam is moreover characterized by modulated crossed fields, the inhomogeneities featuring crossed electric and magnetic fields are of most direct relevance with respect to an actual experimental realization.
In particular, linearly polarized Gaussian beams correspond to fixed orientations $\hat{\vec E}$ and $\hat{\vec B}$.
The experimental realization of purely magnetic fields by means of high intensity lasers requires additional efforts like, e.g., the superposition of two counter propagating laser beams to mutually cancel the electric field components at the beam waist.

To maximize the 
intensity, the focus area is preferably rendered as small as possible.
However, in generic high-intensity laser experiments the focal spot area cannot be chosen at will, but is limited by diffraction.
Assuming Gaussian beams, the effective focus area is conventionally defined to contain $86\%$ of the beam energy ($1/e^2$ criterion for the intensity).
The minimum value of the beam diameter $w_{\rm x}$ in the focus is given by twice the laser wavelength multiplied with $f^\#$,
the so-called $f$-number, defined as the ratio of the focal length and the diameter of the focusing aperture \cite{Siegman};
$f$-numbers as low as $f^\#=1$ can be realized experimentally.
Thus, assuming the pump laser (energy $W_{\rm pump}$, wavelength $\lambda_{\rm pump}$, pulse duration $\tau_{\rm pump}$) to be focused down to the diffraction limit, i.e., $w_{\rm x}=2\lambda_{\rm pump}$, the attainable field strengths are of the order of
\begin{equation}
 {\cal E}^2=I_{\rm pump}\approx \frac{0.86\,W_{\rm pump}}{\tau_{\rm pump}\,\sigma_{\rm pump}}\,,
\label{eq:EBpump}
\end{equation}
with focal area $\sigma_{\rm pump}\approx\pi\lambda_{\rm pump}^2$.
The typical scale of variation associated with the longitudinal evolution of the pump laser beam, i.e., the temporal variation as well as the spatial variation along the pump's propagation direction, is set by the laser wavelength $\lambda_{\rm pump}$.
Hence, the typical temporal variation is clearly of the same order as the typical spatial scales characterizing the field inhomogeneity generated in the focal spot of the pump laser.
Therefore, accounting for the temporal variation is crucial, and a static approximation cannot be justified.

In the present study, we do however not aim at a realistic modelling
of the pump laser, requiring in particular a closer look at the
longitudinal pulse shape evolution. Instead, we rather intend to
account for the most relevant features in the
vicinity of the beam waist where the field strengths are maximal,
while on the other hand keeping the resulting expressions as simple
and compact as possible.{ Therefore, we focus
  on the following $2+1$ dimensional amplitude profile
  (cf. Sec.~\ref{seq:c2+1}),
\begin{equation}
 {\cal E}({\rm x},{\rm y},t)={\cal E}({\rm x})\cos\bigl(\Omega({\rm y}-t)\bigr)\,, \quad \text{where} \quad {\cal E}({\rm x})={\cal E}\,{\rm e}^{-(2{\rm x}/w_{\rm x})^2},
\label{eq:prof2}
\end{equation}
for both cases~\eqref{eq:spec}, to mimic the field inhomogeneity generated in the focal spot of a linearly polarized pump laser (cf. Fig.~\ref{fig:laser}).
This field profile accounts for the plane-wave character of the longitudinal submodulation of the pump laser field.

With respect to an actual experimental realization the field profile in \Eqref{eq:prof2} involves several limitations:
First, we neglect the effect of diffraction spreading about the beam waist, which we expect to constitute a viable approximation within the Rayleigh range $z_{\rm R}$ of the pump laser; in the diffraction limit we have $z_{\rm R}=\pi\lambda_{\rm pump}$ \cite{Siegman}.
Second, we do not account for finite pulse length and longitudinal focusing effects in the field amplitude profile, assuming that these effects amount to sub-leading corrections within the Rayleigh range and for sufficiently long pulse durations $\tau_{\rm pump}$.

The transversal amplitude profile ${\cal E}({\rm x})$ in \Eqref{eq:prof2} squared is easily Fourier transformed to momentum space, 
such that all residual integrations in \Eqref{eq:2+1res} can be performed straightforwardly, resulting in
\begin{multline}
 R_{p,2n}^{(\pm)}=\Theta\bigl(\tfrac{1+\sin\beta}{2}-\tfrac{2n\Omega}{\omega_{\rm in}}\bigr)\,\frac{c_p^2(1+\delta_{n0})^2}{259200\,\pi}\,\alpha^2\left(\frac{e{\cal E}}{m^2}\right)^{4}\left(\frac{\omega_{\rm in}w_{\rm x}}{2}\right)^2\frac{(1-\sin\beta)^4}{(1-\frac{2n\Omega}{\omega_{\rm in}})^2-(\sin\beta-\frac{2n\Omega}{\omega_{\rm in}})^2} \\
 \times{\rm e}^{-\frac{1}{4}\left(\frac{\omega_{\rm in}w_x}{2}\right)^2\left(\cos\beta\pm\sqrt{(1-\frac{2n\Omega}{\omega_{\rm in}})^2-(\sin\beta-\frac{2n\Omega}{\omega_{\rm in}})^2}\right)^2}+{\cal O}\bigl((\tfrac{e{\cal E}}{m^2})^6\bigr)\,,
 \label{eq:res2}
\end{multline}
with $p\in\{1,2\}$, $n\in\{0,\pm1\}$, $c_1=7$ and $c_2=4$.
The number of available photons of frequency $\omega_{\rm in}=2\pi/\lambda_{\rm probe}$ for probing can be estimated from the probe laser energy $W_{\rm probe}$, $N_{\rm probe}\approx{W_{\rm probe}}/{\omega_{\rm in}}$.
The number of induced photons {\it per shot} with particular reflection or transmission kinematics is estimated as $N^{\rm out}=R^{(\pm)}_{p,2n} f_{\rm int} N_{\rm probe}$,
where we have introduced a factor $f_{\rm int}={\rm
  min}\{1,\frac{\tau_{\rm pump}}{\tau_{\rm probe}}\}$, providing a
first estimate of the fraction of the number of incident probe photons
interacting with the inhomogeneity.

As in~\cite{Gies:2013yxa} and \cite{Gies:2014jia} we exemplarily adopt the design parameters of the two high-intensity laser systems to become available in Jena \cite{Jena}:
The terawatt-class laser system JETI~200 \cite{JETI200} ($\lambda_{\rm probe}=800{\rm nm}\approx4.06{\rm eV}^{-1}$, $W_{\rm probe}=4{\rm J}\approx2.50\cdot10^{19}{\rm eV}$,
$\tau_{\rm probe}=20{\rm fs}\approx30.4{\rm eV}^{-1}$) as probe, and the petawatt-class laser system POLARIS \cite{POLARIS} ($\lambda_{\rm pump}=1030{\rm nm}\approx5.22{\rm eV}^{-1}$, $W_{\rm pump}=150{\rm J}\approx9.36\cdot10^{20}{\rm eV}$, $\tau_{\rm pump}=150{\rm fs}\approx228{\rm eV}^{-1}$) as pump.
Hence, the frequency scale $\Omega$ governing the submodulation in \Eqref{eq:prof2} is to be identified with $\Omega = {2\pi}/{\lambda_{\rm pump}}$,  and $f_{\rm int}=1$.
We will specialize to these parameters in the remainder of this paper. 

As discussed in Sec.~\ref{seq:c2+1} above, additional requirements have to be met in order to allow for nonvanishing contributions for $n=1$ in \Eqref{eq:res2}. Namely, we have to ensure that $\frac{2\Omega}{\omega_{\rm in}}<1$ and $\beta\geq\arcsin(\frac{4\Omega}{\omega_{\rm in}}-1)$.
For the JETI~200-POLARIS set-up outlined above, these relations cannot be fulfilled, such that the $n=1$ channel does not give rise to nonzero contributions and $R^{(\pm)}_{p,2}=0$.
Figure~\ref{fig:angles} depicts the emission directions of the induced photons, to be characterized by the angle $\beta'\in(-180^\circ\ldots180^\circ]$ (cf. Fig.~\ref{fig:laser}), as a function of the incidence angle $\beta\in(-180^\circ\ldots180^\circ]$:
As detailed above, apart from the elastic channel without energy exchange between probe photons and pump beam (straight lines labeled by $n=0$), we encounter an inelastic channel (curved lines; $n=-1$).
In the first case, the outgoing directions are fully determined by the kinematics of the probe photons.
Conversely, in the latter case the outgoing directions do not only depend on the probe photon kinematics, but also on the frequency $\Omega$ of the pump beam; cf. Eqs.~\eqref{eq:A2+1res} and \eqref{eq:2+1res}.

In Figs.~\ref{fig:EperpB} and \ref{fig:ausschnitt} we plot the number $N^{\rm out}_p$ of induced photons as a function of the incidence angle $\beta$. The corresponding outgoing angles can be read off from Fig.~\ref{fig:angles}.
Highest numbers of induced photons are achieved for the elastic channel $(n=0)$, while the contributions in the nonvanishing inelastic channel $(n=-1)$ are considerably smaller.
For $\beta=+90^\circ$ the induced contributions vanish for all channels,
reflecting the well known fact that the invariant $(sk)|_{k^2=0}=\vec{k}\cdot(\hat{\vec E}\times\hat{\vec B})-\omega$ vanishes in crossed fields of the same amplitude for this particular choice of $\beta$ (cf. Appendix~\ref{app:conscrossfields}).
Contrarily, for $\beta=-90^\circ$ the induced contributions in the elastic channel diverge; the reason for this unphysical divergence is the idealization of an infinite extent of the inhomogeneity along $\vec{e}_{\rm y}$, allowing for an infinitely long interaction region at ``grazing incidence''.
This divergence is physically cut off by the finite range of any experimentally attainable field inhomogeneity, cf. also Sec.~\ref{seq:Exp}.

However, we still expect to retain the basic features of
Figs.~\ref{fig:EperpB} and \ref{fig:ausschnitt} (left), like a
substantial enhancement of the number of induced photons to $N^{\rm
  out}\gtrsim1$ in the vicinity of $\beta=-90^\circ$, particularly in
the angle range depicted in Fig.~\ref{fig:ausschnitt}. Let us
demonstrate that this is indeed the case.  As the divergence is
encountered in the elastic $n=0$ contribution, we argue that it is
sufficient to show that the induced contributions independent of
$\Omega$ do no longer diverge for $\beta\to-90^\circ$ when the extent
of the inhomogeneity along $\vec{e}_{\rm y}$ is rendered finite.
Correspondingly, for the present discussion we can specialize to
$\Omega=0$ from the outset [cf. also the argument given in the second
  paragraph below \Eqref{eq:Aout1+1magnetic}].
Recall, that the longitudinal field profile of a real Gaussian beam about its beam waist depletes as $\sim1/\sqrt{1+({\rm y}/z_{\rm R})^2}$, with Rayleigh range $z_{\rm R}$ \cite{Siegman}.
In order to demonstrate that the divergence is regularized as soon as the extent of the field inhomogeneity along $\vec{e}_{\rm y}$ finite, we thus focus on the alignments~\eqref{eq:spec} and a static two-dimensional field profile,
\begin{equation}
 {\cal E}({\rm x},{\rm y})=\frac{{\cal E}({\rm x})}{\sqrt{1+({\rm y}/\zeta)^2}}\,, \quad \text{where} \quad {\cal E}({\rm x})={\cal E}\,{\rm e}^{-(2{\rm x}/w_{\rm x})^2}.
\label{eq:prof2a}
\end{equation}
This corresponds to \Eqref{eq:prof2} evaluated at $\Omega=0$ and multiplied with an overall longitudinal damping factor of $1/\sqrt{1+({\rm y}/\zeta)^2}$.
For the choice $\zeta=z_{\rm R}=\pi\lambda_{\rm pump}$
the longitudinal field profile decreases in the same way as a Gaussian beam. Conversely, for $\zeta\to\infty$ the infinitely extended inhomogeneity is recovered.
Plugging the field profile~\eqref{eq:prof2a} into the scattering amplitude~\eqref{eq:f(k,k)_2},
we determine the number of reflected photons 
in the following way: we integrate the differential cross section~\eqref{eq:diffcross} over all values of $\beta'$ which characterize induced photon waves propagating into the backward half-space  
(cf. Fig.~\ref{fig:2Dplane}), and multiply the results by $N_{\rm probe}/w_{\rm probe}$. Assuming also the probe beam to be focused down to the diffraction limit, we identify $w_{\rm probe}=2\lambda_{\rm probe}$.
In Fig.~\ref{fig:ausschnitt} (right) we depict our results for different values of $\zeta=\{1,5,10,50\}z_{\rm R}$.
The emergence of an unphysical divergence at $\beta\to-90^\circ$ when increasing $\zeta$ to large (unphysical) values is clearly visible.
The numbers of induced photons tend to be somewhat larger than in Fig.~\ref{fig:ausschnitt} (left), which can be attributed to the $\beta'$ integration over all backward directions, covering an angular range of $\pi$.
As a function of the incidence angle $\beta$, the number of induced photons in the elastic mode becomes larger than one in roughly the same angular range for all choices of $\zeta$ depicted in Fig.~\ref{fig:ausschnitt} (right), most importantly also for the physically motivated choice of $\zeta=z_{\rm R}$.
A more in depth analysis of the precise behavior about $\beta=-90^\circ$ is outside the scope of our present work, and will be subject of further investigations.

Focusing on the results of the $2+1$ dimensional calculation, for the JETI~200-POLARIS set-up, the number of reflected photons
becomes larger than one for angles in the range between $\beta=-101.79^\circ$ $\leftrightarrow$ $\beta'=-78.21^\circ$ 
and $\beta=-78.21^\circ$ $\leftrightarrow$ $\beta'=-101.79^\circ$ (cf. Fig.~\ref{fig:angles}), i.e., both incident and outgoing angles within $\approx \pm10^\circ$ about $-\vec{e}_{\rm y}$.
This should be compared to the divergence of a Gaussian beam focused down to the diffraction limit, $\theta\approx\frac{1}{\pi}\approx18^\circ$.
This makes the detection of reflected photons within $10^\circ$ about the beam axis a nontrivial experimental issue.
In the present ``two-color'' JETI~200-POLARIS set-up, we expect that frequency filtering techniques can be used for sufficient background suppression. In a single laser set-up using a suitably split beam, background suppression may require special pulse design and a high time resolution for the probe photon detection.
Finally, let us emphasize that our estimates presented here are of the
same order of magnitude and thus compatible with our previous --
much less realistic -- estimates based on a static one-dimensional,
purely magnetic field inhomogeneity \cite{Gies:2013yxa}.

\begin{figure}[h]
\center
\includegraphics[width=0.47\textwidth]{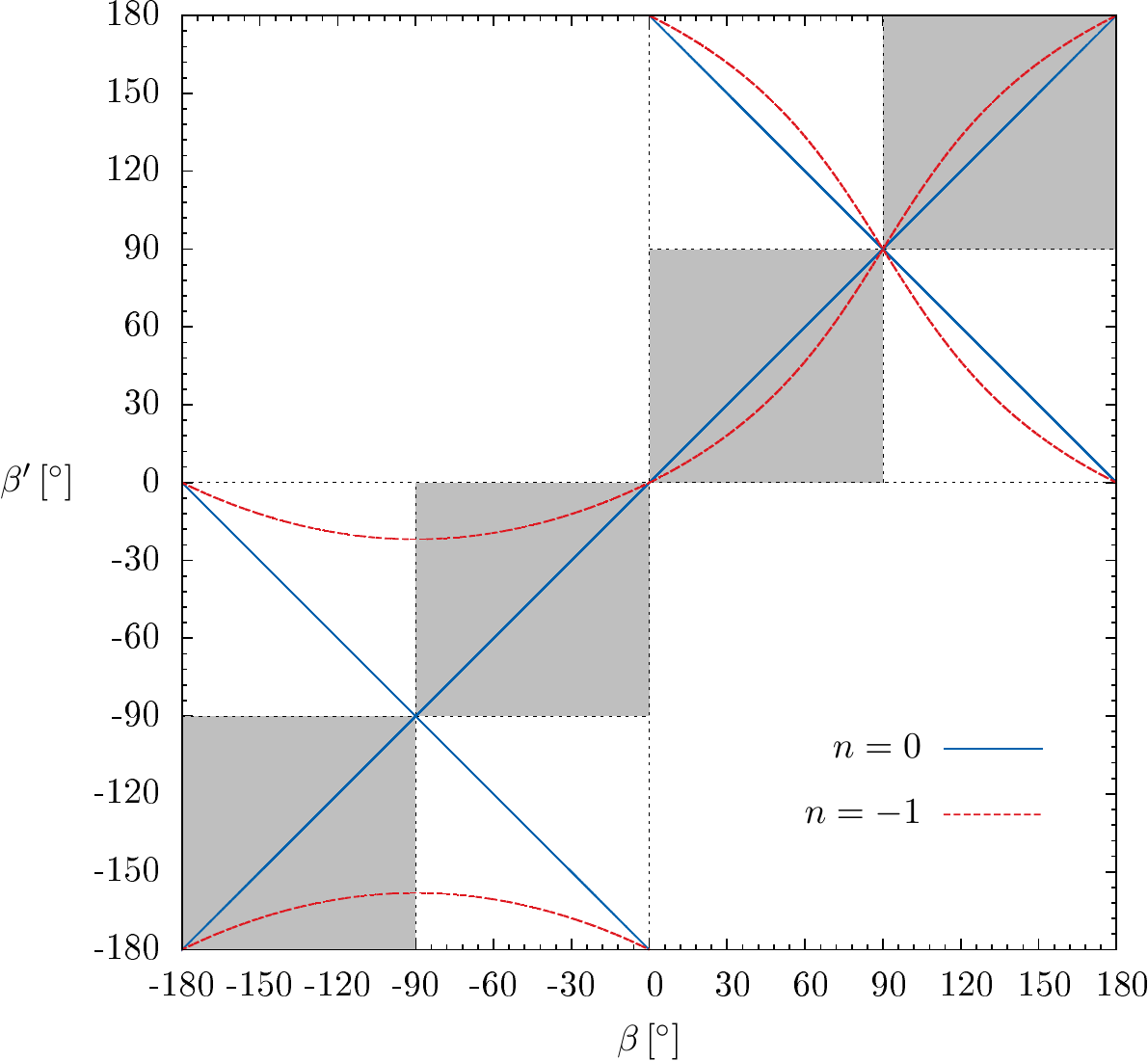}
\caption{Plot of the outgoing angle $\beta'=\varangle(\vec{k}',\vec{e}_{\rm x})$ as a function of the incidence angle $\beta=\varangle(\vec{k}_{\rm in},\vec{e}_{\rm x})$, adopting the design parameters of the Jena high-intensity laser systems, JETI~200 and Polaris (cf. main text) in the $n=0$ and $n=-1$ channels; no photons are induced in the $n=1$ channel for this set-up. Quantum reflection gives rise to induced photons propagating in ``forward'' (gray-shaded) and ``backward'' (white) directions;
cf. also Fig.~\ref{fig:2dim}.}
\label{fig:angles}
\end{figure}

\begin{figure}[h]
\center
\subfigure{\includegraphics[width=0.5\textwidth]{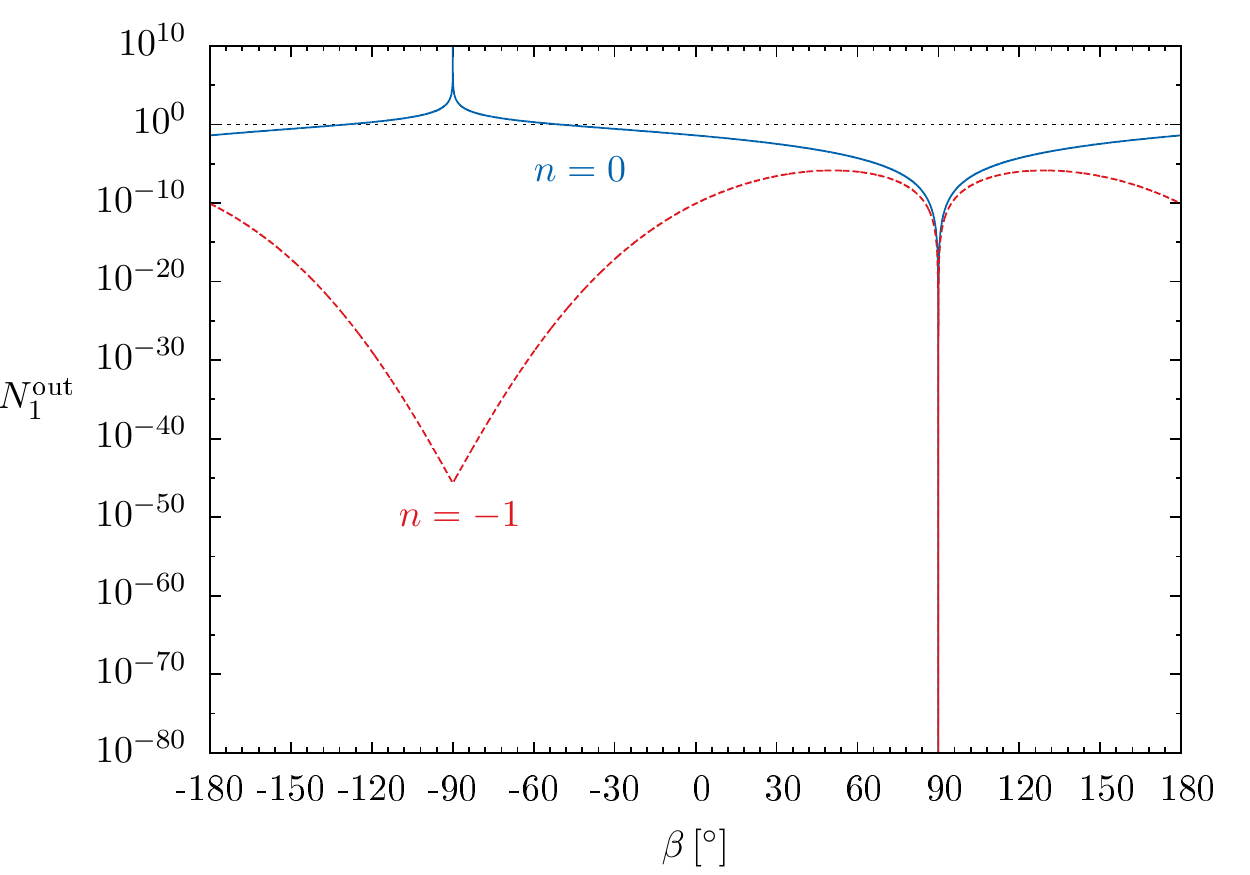}}\hspace*{0.1cm}
\subfigure{\includegraphics[width=0.5\textwidth]{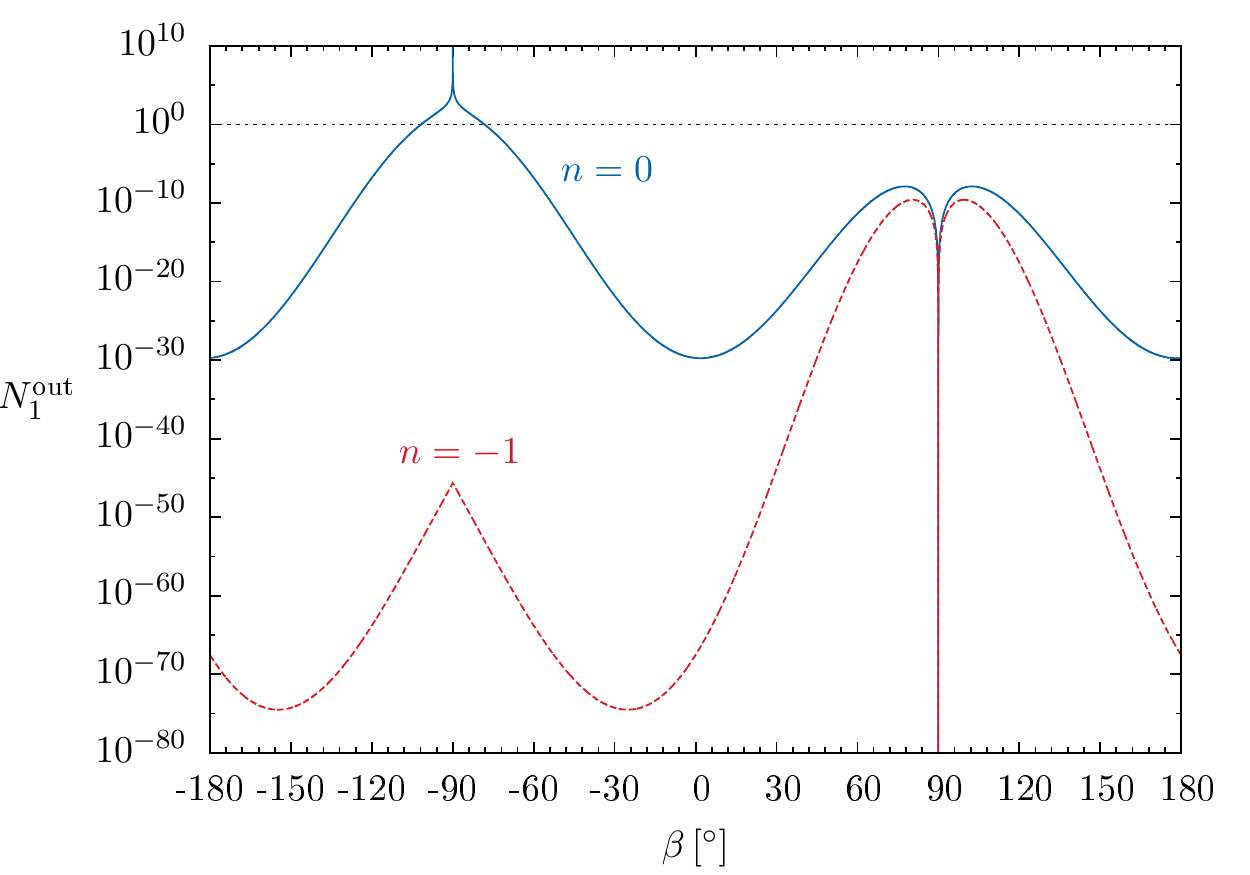}}
\caption{Number of induced photons for the field inhomogeneity~\eqref{eq:prof2}, $p=1$ and $n\in\{-1,0\}$ plotted as a function of the incidence angle $\beta$ for the JETI~200 and POLARIS set-up detailed in the main text; no photons with $n=1$ are induced.
For $\beta=90^\circ$ all contributions vanish.
Conversely, for $\beta=-90^\circ$ the number of induced photons with $n=0$ diverges (cf. main text).
The corresponding outgoing angles can be inferred from Fig.~\ref{fig:angles}. The results for $p=2$ follow by multiplication with a factor of $(c_2/c_1)^2\approx0.33$. {\bf Left:} Forward direction. While the $n=0$ contribution is induced in probe laser direction, $n\neq0$ contributions are deflected. {\bf Right:} Backward direction.}
\label{fig:EperpB}
\end{figure}

\begin{figure}[h]
\center
\subfigure{\includegraphics[width=0.5\textwidth]{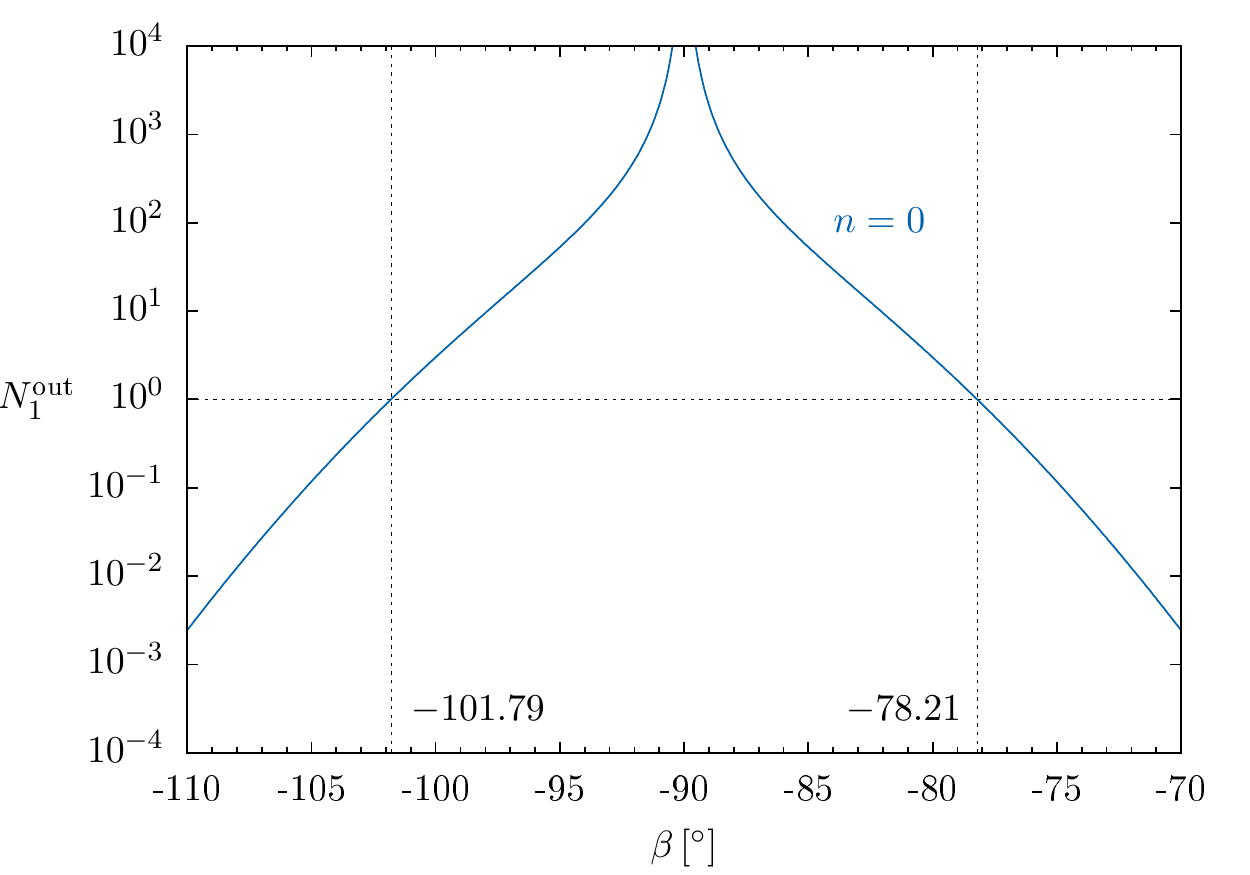}}\hspace*{0.1cm}
\subfigure{\includegraphics[width=0.5\textwidth]{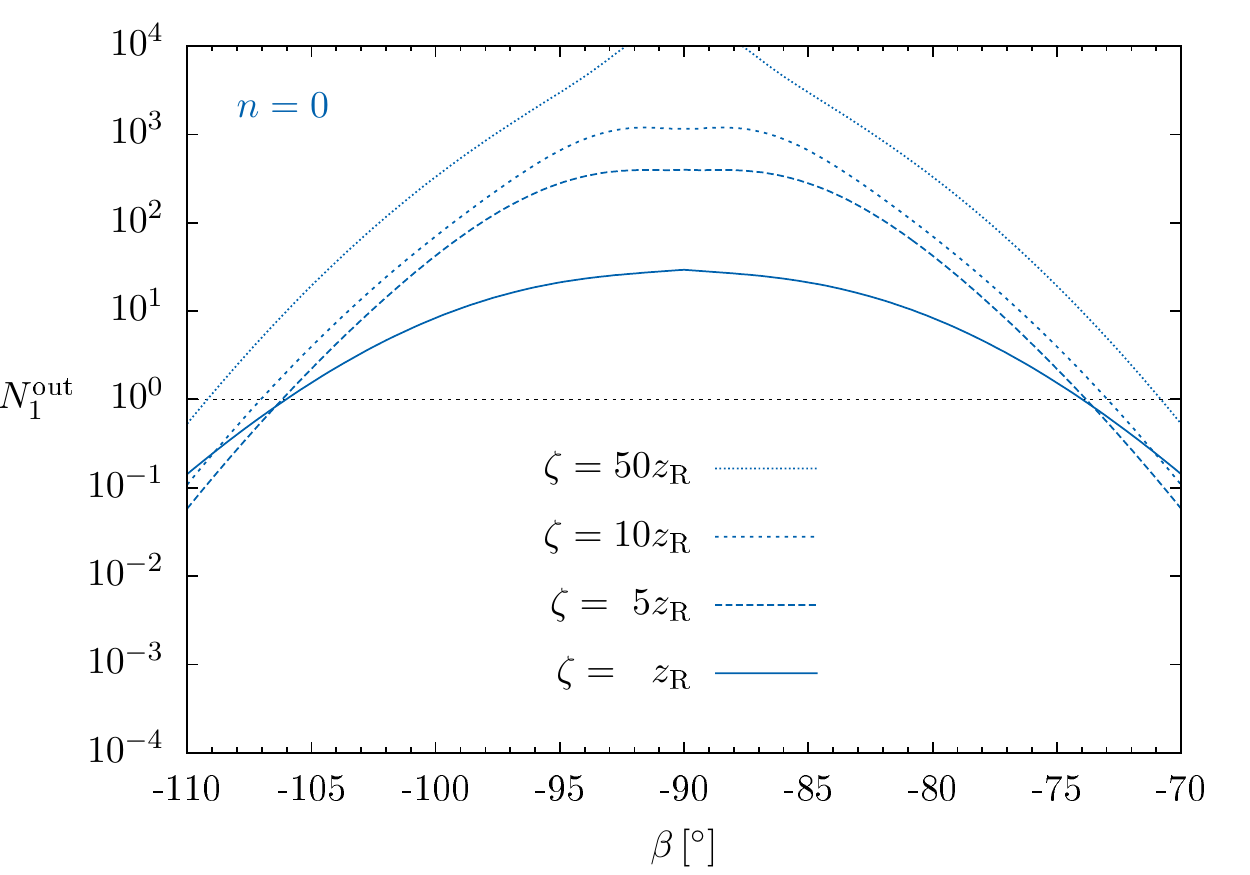}}
\caption{Number of induced photons in backward direction in the
  angular range where $N_{p=1}^{\rm out}\gtrsim1$. {\bf Left:} Close
  up of Fig.~\ref{fig:EperpB} (right) for the 2+1 dimensional
  inhomogeneity~\eqref{eq:prof2}.  {\bf Right:} Static two dimensional
  inhomogeneity~\eqref{eq:prof2a}. For this result, we integrated over
  all photons induced in backward direction for elastic
  kinematics (cf. main text). The divergence at $\beta=-90^\circ$ is
  physically regularized by the finite extent of the inhomogeneity
  along $\vec{e}_{\rm y}$ specified here in multiples of the Rayleigh
  range $z_{\text{R}}$.}
\label{fig:ausschnitt}
\end{figure}

\section{Conclusions} \label{seq:Con}

In this paper we have investigated the phenomenon of quantum reflection of probe photons off a strong inhomogeneous field for a large class of field inhomogeneities.
This includes spatio-temporal field inhomogeneities with up to two nontrivial spatial directions facilitating the exploration of this phenomenon on a new level.
In particular the time-dependence gives rise to new effects such as frequency mixing though on a quantitatively sub-dominant level.
In general, our new results -- though much more detailed and realistic -- support our earlier estimates based on one-dimensional static inhomogeneities \cite{Gies:2013yxa}.
This endorses our viewpoint that quantum reflection can become a valuable novel means to probe the nonlinearity of the quantum vacuum.
A particularly beneficial feature is given by the fact that a signal-to-background separation is supported by the kinematics of searching for reflected photons in the field free region.

Our approach as already suggested in \cite{Gies:2013yxa} heavily
relies on analytical insights into the photon polarization tensor in
constant and homogeneous electromagnetic field backgrounds. Resorting
to a given electromagnetic field alignment the inhomogeneity is built
in {\it a posteriori} by means of the locally constant field
approximation.  In the present paper we have discussed both purely
magnetic field and crossed-field inhomogeneities. For the application
of our approach, it is important to realize, that the Ward identity
can in fact be violated within this approximation scheme. However, we
have been able to identify those kinematic and polarization
configurations for which the Ward identity can be guaranteed to hold
exactly even within the approximation scheme. The resulting
requirements include $2+1$ dimensional inhomogeneities compatible with
field configurations attainable in realistic high-intensity laser
experiments.  We emphasize, however, that the present approach
if applied to generic $3+1$ dimensional field inhomogeneities
would need new theoretical insights that go beyond the
locally constant field approximation.

Building on contemporary knowledge about the photon polarization
tensor, we present fully analytical estimates for the effect of
quantum reflection in a pump-probe type set-up consisting of a typical
terawatt- and petawatt-class high-intensity laser system.  In support
of our previous results \cite{Gies:2013yxa}, our new estimates are
based on much more realistic, spatio-temporal electromagnetic field
configurations, compatible with the inhomogeneous fields attainable in
the focal spots of high-intensity laser systems.  The fact that we
obtain similar numbers of induced photons experiencing quantum
reflection, corroborates the robustness of the effect with respect to
generalizations of the inhomogeneity.
Correspondingly, our investigations establish quantum reflection as a prospective candidate for the experimental verification of quantum vacuum nonlinearity in strong electromagnetic fields.

\section*{Acknowledgments}

The authors are grateful to Matt~Zepf for many interesting, fruitful
and very enlightening discussions. HG acknowledges support by the DFG
under grants Gi 328/5-2 (Heisenberg program) and SFB-TR18.

\appendix

\section{Photon polarization tensor in the weak-field limit}\label{sec:appendix}

\subsection{Magnetic background field}\label{app:B=const.}

On the light cone $k^2=0$, the photon polarization tensor in a constant homogeneous magnetic field takes a particularly simple form.
In the perturbative weak-field limit it is conveniently represented as (see \cite{Dittrich:2000zu,Karbstein:2013ufa} and references therein)
\begin{equation}
 \Pi^{\mu\nu}(k)\big|_{k^2=0}=\Pi^{\mu\nu}_{(2)}(k)(eB)^{2}+{\cal O}\bigl((eB)^4\bigr)\,,
\end{equation}
with tensorial coefficient
\begin{equation}
 \Pi^{\mu\nu}_{(2)}(k)= \frac{1}{45} \frac{\alpha}{\pi} \Bigl(7\,k_\parallel^2\,P_\parallel^{\mu\nu}(k)-4\,k_\perp^2\,P_\perp^{\mu\nu}(k)\Bigr)\frac{1}{m^4}\,, \label{eq:pi2olc}
\end{equation}
and projectors onto photon polarization modes defined in \Eqref{eq:Proj}.
Here we expressed the tensorial coefficient~\eqref{eq:pi2olc} in a particularly suggestive form, making use of $k^2=0\ \leftrightarrow\ k_\parallel^2=-k_\perp^2$ for the scalar prefactors of the projectors $P_p^{\mu\nu}(k)$. 
In this particular limit, the tensorial quantity $\pi^{\mu\nu}_{(2)}(k,k')$ as defined in \Eqref{eq:pi} is obtained straightforwardly.
For a purely magnetic background, it is given by
\begin{equation}
 \pi^{\mu\nu}_{(2)}(k,k')=  \frac{1}{90} \frac{\alpha}{\pi} \, \biggl[7\Bigl(k_\parallel^2\,P_\parallel^{\mu\nu}(k)+k_\parallel'^2\,P_\parallel^{\mu\nu}(k')\Bigr)
  -4\Bigl(k_\perp^2\,P_\perp^{\mu\nu}(k)+k_\perp'^2\,P_\perp^{\mu\nu}(k')\Bigr)\biggr]\frac{1}{m^4}\,.
\end{equation}
For the two cases specified in \Eqref{eq:Bcases}, a contraction with the corresponding global projector $\tilde P^{\mu\nu}_p$  results in
\begin{equation}
 (i):\quad\pi_{\parallel,(2)}=\frac{7}{45} \frac{\alpha}{\pi} \, k_\parallel^2\,\frac{1}{m^4}\,, \quad\text{and}\quad (ii):\quad\pi_{\perp,(2)}=-\frac{4}{45} \frac{\alpha}{\pi} \, k_\perp^2\,\frac{1}{m^4} \,.
\end{equation}

\subsection{Constant crossed fields} \label{app:conscrossfields}

Analogous statements hold for the  photon polarization tensor in constant crossed fields,
characterized by $\vec{E} \cdot \vec{B} = 0$ and ${\cal E}=|\vec{E}| =|\vec{B}|$, i.e., perpendicular electric and magnetic fields of the same magnitude \cite{narozhnyi:1968}.
In this case it is convenient to introduce the four vector $s^\mu=(1,\hat{\vec{E}}\times\hat{\vec{B}})$.
In the weak-field limit and for $k^2=0$ the photon polarization tensor is then given by (cf. also \cite{Dittrich:2000zu})
\begin{equation}
 \Pi^{\mu\nu}(k)\big|_{k^2=0}=\Pi^{\mu\nu}_{(2)}(k)(e{\cal E})^{2}+{\cal O}\bigl((e{\cal E})^4\bigr)\,,
\end{equation}
with tensorial coefficient
\begin{equation}
 \Pi^{\mu\nu}_{(2)}(k)= - \frac{1}{45} \frac{\alpha}{\pi} (sk)^2 \, \Bigl(7\,P_1^{\mu\nu}(k)+4\,P_2^{\mu\nu}(k)\Bigr)\frac{1}{m^4}\,,
\end{equation}
and projectors onto photon polarization modes defined as
\begin{equation}
 P_1^{\mu\nu}(k)=\frac{(^*Fk)^\mu(^*Fk)^\nu}{{\cal E}^2(sk)^2}\,, \quad P_2^{\mu\nu}(k)=\frac{(Fk)^\mu (Fk)^\nu}{{\cal E}^2(sk)^2}\,. \label{eq:P_12}
\end{equation}
Here $^*F^{\mu\nu}=\frac{1}{2}\epsilon^{\mu\nu\alpha\beta}F_{\alpha\beta}$ denotes the dual field strength tensor, and we have made use of the shorthand notation $(sk)=s^\alpha k_\alpha$, $(^*Fk)^\mu=^*\!\!F^{\mu\alpha}k_\alpha$ and $(Fk)^\mu=F^{\mu\alpha}k_\alpha$.
Correspondingly, we obtain [cf. \Eqref{eq:pi}]
\begin{equation}
 \pi^{\mu\nu}_{(2)}(k,k')=  -\frac{1}{90} \frac{\alpha}{\pi} \, \biggl[7\Bigl((sk)^2\,P_1^{\mu\nu}(k)+(sk')^2\,P_1^{\mu\nu}(k')\Bigr)
  +4\Bigl((sk)^2\,P_2^{\mu\nu}(k)+(sk')^2\,P_2^{\mu\nu}(k')\Bigr)\biggr]\frac{1}{m^4}\,.
\end{equation}

For this particular field configuration it is convenient to make use of a more explicit representation of the projectors:
Without loss of generality we set $\vec{s}=\vec{e}_{\rm y}$ and write the directions of the magnetic and electric fields as $\hat{\vec{B}}=(\cos\varphi,0,\sin\varphi)$ and $\hat{\vec{E}}=(-\sin\varphi,0,\cos\varphi)$, respectively.
The angle $\varphi\in[0\ldots2\pi)$ parameterizes all possible orientations of the electric and magnetic field vectors for constant crossed fields fulfilling $\vec{s}=\vec{e}_{\rm y}$.
The projectors~\eqref{eq:P_12} can then be represented as $P_1^{\mu\nu}=u_1^\mu u_1^\nu$ and $P_2^{\mu\nu}=u_2^\mu u_2^\nu$ with
\begin{align}
 u_1^\mu&=\bigl(\tfrac{k_{\rm z}\sin\varphi+k_{\rm x}\cos\varphi}{(sk)},-\cos\varphi,\tfrac{k_{\rm z}\sin\varphi+k_{\rm x}\cos\varphi}{(sk)},-\sin\varphi\bigr)\,, \nonumber\\
 u_2^\mu&=\bigl(\tfrac{k_{\rm z}\cos\varphi-k_{\rm x}\sin\varphi}{(sk)},\sin\varphi,\tfrac{k_{\rm z}\cos\varphi-k_{\rm x}\sin\varphi}{(sk)},-\cos\varphi\bigr)\,. \label{eq:us}
\end{align}
It is easy to see that for general kinematic situations $u_1^\mu$ is invariant under $\vec{k}\to \vec{k}+ c(k) \hat{\vec{E}}$,
while $u_2^\mu$ is invariant under $\vec{k}\to \vec{k}+ c(k) \hat{\vec{B}}$,
where $c(k)$ is an arbitrary function of the components of $k^\mu$.

For the special situation where $k_{\rm z}\equiv0$, we can identify two cases where either $u_1^\mu$ or $u_2^\mu$ assumes a particularly simple form:
For $\varphi=\frac{\pi}{2} $ and $k_{\rm z}=0$ we have $u_1^\mu=(0,-\vec{e}_{\rm z})$, while  for $\varphi=0$ and $k_{\rm z}=0$ we have $u_2^\mu=(0,-\vec{e}_{\rm z})$.
Obviously, for these special choices the projectors $P_1^{\mu\nu}$ or $P_2^{\mu\nu}$, respectively, are independent of the values of $k_{\rm x}$, $k_{\rm y}$ and $\omega$, i.e., are conserved for general kinematics restricted to the ${\rm x}$-${\rm y}$ plane.  

For the cases considered in Sec.~\ref{seq:Cross} a contraction with the corresponding global projector $\tilde P^{\mu\nu}_p$ yields
\begin{equation}
 \pi_{1,(2)}(k,k')=-\frac{7}{45} \frac{\alpha}{\pi} \, \frac{(sk)^2+(sk')^2}{2}\,\frac{1}{m^4}\,, \quad\text{and}\quad \pi_{2,(2)}(k,k')=-\frac{4}{45} \frac{\alpha}{\pi} \, \frac{(sk)^2+(sk')^2}{2}\,\frac{1}{m^4} \,.
\end{equation}

\end{document}